\documentclass[PAPER, cernpreprint, atlasdraft=false, UKenglish, texmf, orcidlogo]{atlasdoc}
\usepackage{atlaspackage}
\usepackage{atlasbiblatex}

\usepackage{atlasphysics}

\usepackage{amsmath}
\usepackage{rotating}
\usepackage{graphicx}
\usepackage{multirow}

\newcolumntype{~}{>{\global\let\currentrowstyle\relax}}
\newcolumntype{^}{>{\currentrowstyle}}

\mathchardef\mhyphen="2D

\graphicspath{{logos/}{figures/}}

\usepackage{ANA-EXOT-2022-27-PAPER-defs}

\AtlasTitle{Search for electroweak production of vector-like leptons in $\tau$-lepton and $b$-jet final states in $\ensuremath{pp}$
collisions at $\sqrt{s}$ = 13 TeV with the ATLAS detector}

\AtlasAbstract{%
A search for pair-production of vector-like leptons is presented, considering their decays into a third-generation Standard Model (SM) quark and a vector leptoquark ($U_1$) as predicted by an ultraviolet-complete extension of the SM, referred to as the `4321' model. Given the assumed decay of $U_1$ into third-generation SM fermions, the final state can contain multiple $\tau$-leptons and $b$-quarks. This search is based on a dataset of $pp$ collisions at $\sqrt{s}=13$~TeV recorded with the ATLAS detector during Run 2 of the Large Hadron Collider, corresponding to an integrated luminosity of up to $\ensuremath{140~\mathrm{fb}^{-1}}\xspace$. %
No significant excess above the SM background prediction is observed, and 95\% confidence level limits on the cross-section times branching ratio are derived as a function of the vector-like lepton mass. A lower observed (expected) limit of 910 GeV (970 GeV) is set on the vector-like lepton mass. Additionally, the results are interpreted for a supersymmetric model with an $R$-parity violating coupling to the third-generation quarks and leptons. Lower observed (expected) limits are obtained on the higgsino mass at 880 GeV (940 GeV) and on the wino mass at 1170 GeV (1170 GeV).

}

\AtlasRefCode{EXOT-2022-27}

\PreprintIdNumber{CERN-EP-2025-055}

\AtlasJournal{EPJC}
\AtlasJournalRef{Eur. Phys. J. C 85 (2025) 1335}
\AtlasDOI{10.1140/epjc/s10052-025-14748-z}

\usepackage{authblk}
\AtlasCoverEgroupAnalysisTeam{atlas-exot-2022-27-analysis-team@cern.ch}

\hypersetup{pdftitle={ATLAS document},pdfauthor={The ATLAS Collaboration}}

\begin{document}

\maketitle

\section{Introduction}
%

%
The Standard Model (SM) adequately explains sub-atomic particles and their interactions at the sub-TeV scale. However, a number of phenomena require additional understanding, such as the fine-tuning problem of the Higgs boson mass, the mass hierarchy of the SM particles, the similarities in the lepton and quark flavour generations, the nature of dark matter in the universe, the origin of neutrino oscillations; a selection of them has been discussed in Refs.~\cite{Susskind79,DimopolousGeorgi81,WittenGoodman85,ParticleDataGroup:2024pth}. %
In the quest to give a coherent explanation of the experimental results which point to deviations from the SM expectation in semileptonic $B$-meson decays, denoted as `flavour anomalies'
\cite{BaBar:2012obs,BaBar:2013mob,Belle:2019rba,LHCb:2020gog,LHCb:2020lmf,LHCb:2017smo,LHCb:2023zxo}
, a number of beyond-the-SM (BSM) theories are constructed. The minimal requirements for a unified explanation of all anomalies are new physics contributions in neutral currents ($b \rightarrow s \mu \mu$ ~\cite{LHCb:2022qnv}) and charged currents (enhanced decay rates of $b \rightarrow c \tau \nu$ ~\cite{Huang:2018nnq}), as argued in Ref.~\cite{DiLuzio:2018zxy}.
At the same time, the BSM theory also needs to be consistent with the non-observation of related signals in other flavour-changing processes \cite{Alonso:2016oyd}, electroweak (EW) precision observables, $\tau$-lepton decays \cite{Feruglio:2016gvd,Feruglio:2017rjo} and high transverse momentum ($\pt$) searches \cite{Faroughy:2016osc,Greljo:2017vvb}. %

Typical attempts for the explanation of flavour anomalies often lack consideration of additional observations, like in Refs.~\cite{Bhattacharya:2014wla,Alonso:2015sja,Greljo:2015mma,Calibbi:2015kma,Bauer:2015knc,Fajfer:2015ycq,Becirevic:2016yqi,Buttazzo:2017ixm,Marzocca:2018wcf,Azatov:2018knx,Kumar:2018kmr}.
Constraints from low-energy precision measurements can be satisfied when considering mainly left-handed semi-leptonic currents and flavour symmetry with dominant couplings to third generation fermions.
This can be achieved by supplementing the SM with a single field - a vector leptoquark \Umu, where leptoquarks, through their couplings to a quark and a lepton, induce semi-leptonic transitions of $B$-mesons at tree level. In order to have an ultra-violet (UV) complete theory, massive vectors require either compositeness (e.g. composite vector resonance \cite{Barbieri:2015yvd,Buttazzo:2016kid,Barbieri:2016las}) or a spontaneously broken gauge theory.
The second approach~\cite{DiLuzio:2017vat} of embedding the vector leptoquark \Umu into a spontaneously broken gauge theory is considered in the search presented in this paper.

The gauge structure of the considered model is invariant under the local group $\mathcal{G}_{4321} \equiv SU(4) ~\times~ SU(3)'~ \times SU(2)_L ~\times~ U(1)'$ \cite{DiLuzio:2018zxy}, denoted as the \vll model, which is fully calculable at the UV energy scale.
It provides an explanation of the `flavour anomalies' by allowing sizable (maximal) flavour violation in quark-lepton currents, where the neutral current anomaly can be well accommodated in the model. Consequently, the large leptoquark flavour violating couplings trigger new physics contributions in charged currents.
It is also compatible with additional observations, such as the absence of any significant excess in searches for leptoquarks~\cite{EXOT-2019-13,EXOT-2021-15,SUSY-2018-12,EXOT-2019-15,SUSY-2018-34,EXOT-2020-08}, and the non-observation of flavour anomalies in (first or second generation) purely hadronic and purely leptonic channels \cite{DiLuzio:2017vat,Bordone:2017bld,DiLuzio:2018zxy,Cornella:2021sby}.
The symmetry breaking leads to the appearance of a TeV-scale vector leptoquark with quantum numbers favoured by the flavour anomalies~\cite{ParticleDataGroup:2024pth}. %

The particle content of the model has three heavy gauge bosons: a colour octet (\gp), a weak isospin-singlet vector leptoquark (\Umu, electric charge 2/3\e), and a colour singlet \Zp boson. Additionally, there are generations of vector-like fermions: vector-like quarks (VLQ) $U/D, C/S, T/B$ and vector-like leptons (VLL) $N_1/E_1, N_2/E_2, N_3/E_3$. The VLLs in this model are set to be SU(2) doublets with a charged and a neutral component. Only one generation of VLLs is present in the model, and it is assumed to be fully mass degenerate.
The VLLs are hypothetical, non-chiral, colour-singlet, spin-1/2 particles predicted in various BSM scenarios, primarily aimed at addressing the flavour hierarchy in Higgs boson couplings and the fine-tuning problem associated with the Higgs mass. They can be electrically neutral or carry a charge of $\pm$1\e, can decay to the SM leptons via direct mixing or via other BSM portals while conserving their electric charges and lepton numbers, and hence are classified by the SM lepton generation number.
This search targets the pair production of VLLs in proton--proton ($pp$) collisions at $\sqrt{s}$ = 13 TeV using the data collected by the ATLAS detector during Run 2 of the Large Hadron Collider (LHC). A previous search for VLLs in the \vll model was performed by the CMS Collaboration, where a local $2.8 \sigma$ excess was reported using the Run-2 data~\cite{CMS-B2G-21-004} at a representative VLL mass, \mVLL, of 600 GeV, while somewhat similar or milder excesses are present for other \mVLL values in the range of 500 -- 1100 GeV.

\begin{figure}[t]
\centering
\subfloat[]{\includegraphics[width=0.32\textwidth]{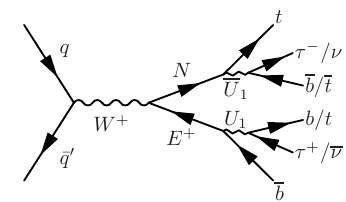}} %
\subfloat[]{\includegraphics[width=0.32\textwidth]{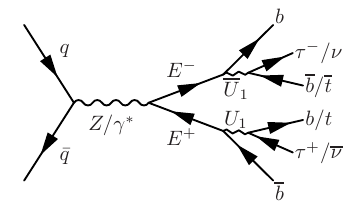}} %
\subfloat[]{\includegraphics[width=0.32\textwidth]{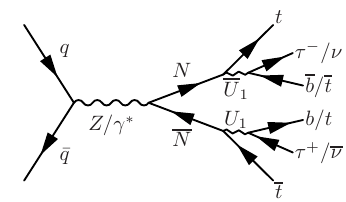}} %
\caption{Vector-like lepton pair production and decays in the \vll model for (a) $pp\rightarrow E^{+}N$, (b) $pp\rightarrow E^+E^-$, (c) $pp\rightarrow N\overline{N}$.}
\label{fig:VLLfeynD}
\end{figure}

The analysis targets the EW production of a pair of VLLs ($E^{+}N$, $E^+E^-$ or $N\overline{N}$) via decays of off-shell $W^{\pm}$ or $Z/\gamma^*$ bosons in the \vll model.
Each VLL decays through $U_1$ as $E^{+} \rightarrow U_{1} \bar{b}$ and $N \rightarrow \overline{U}_{1} t$ followed by $U_{1} \rightarrow t \bar{\nu}_{\tau}$ or $U_{1} \rightarrow b \tau^{+}$, as shown in Figure~\ref{fig:VLLfeynD}.
The branching ratios (BR) for $E$ and $N$ decays are roughly $50\%$ for all processes (BR($E^{+}\rightarrow b\bar{b}\tau^{+}$), BR($E^{+}\rightarrow t\bar{b} \bar{\nu}_{\tau}$), BR($N\rightarrow \bar{b}t\tau^{-}$) and BR($N \rightarrow \bar{t}t \nu_{\tau}$)), because \Umu decays with equal probability into $t\bar{\nu}_{\tau}$ and $b\tau^{+}$ due to the SU(2) gauge symmetry.
The \mVLL is the single free parameter of the model, investigated in the range 200 -- 1500 GeV.

The search is performed probing simultaneously the $E^+E^-$, $E\bar{N}$ and $N\bar{N}$ production, which gives rise to a signature with a large number of jets, of which four contain a $b$-hadron (\bjets), and up to four hadronically decaying $\tau$-leptons (\tauh). The search utilises the full Run-2 $pp$ dataset collected by the ATLAS experiment at a centre-of-mass energy $\sqrt{s}=13$ TeV.  Events are selected using a combination of triggers satisfying selection requirements for light leptons ($\ell$= $e$, $\mu$), \tauh, \bjet, or missing transverse energy (\met). The dataset selected with light leptons, \tauh, or \met triggers corresponds to an integrated luminosity of \lumival, while that selected with \bjet triggers corresponds to \lumivalbjet.

\begin{figure}[t]
\centering
\subfloat[]{\includegraphics[width=0.32\textwidth]{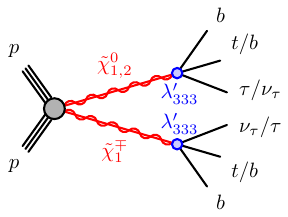}} %
\subfloat[]{\includegraphics[width=0.32\textwidth]{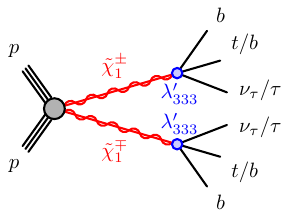}} %
\subfloat[]{\includegraphics[width=0.32\textwidth]{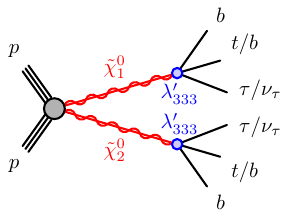}} %
\caption{Diagrams for SUSY $R$-parity violation model with $\lambda'_{333}$ coupling for wino or higgsino production of (a) $pp\rightarrow \tilde\chi^{\pm}_{1} \tilde\chi^{0}_{1,2}$, (b) $pp\rightarrow \tilde\chi^{\pm}_{1} \tilde\chi^{\mp}_{1}$, (c) $pp\rightarrow \tilde\chi^{0}_{1} \tilde\chi^{0}_{2}$.}
\label{fig:SUSYfeynD}
\end{figure}

An additional interpretation is performed for a supersymmetry (SUSY) model~\cite{Golfand:1971iw, Volkov:1973ix, Wess:1974tw, Wess:1974jb, Ferrara:1974pu, Salam:1974ig} with $R$-parity violation (RPV), which can accommodate `flavour anomalies'
and the $(g-2)_{\mu}$ anomaly \cite{Chakraborty:2015bsk}.
It is realized using the lepton-number-violating RPV coupling to third generation quarks and leptons ($LQ\bar{D}$ , where $L$, $Q$ and $\bar{D}$ are the lepton doublet, quark doublet and down-type quark superfields, respectively).  The corresponding interaction term has a coefficient $\lambda'_{333}$, where `$333$' denotes the assumed couplings to the third generation fermions only.
In the minimal SUSY extension of the SM (MSSM) \cite{Fayet:1976et,Fayet:1977yc} the wino, bino and higgsino fields are superpartners of SM gauge boson ($W$, $\gamma/Z$), and Higgs fields, respectively. These states mix to form mass eigenstates at the electroweak scale: the charginos ($\tilde\chi^{\pm}_{1,2}$) and neutralinos ($\tilde\chi^{0}_{1,2,3,4}$), with indices denoting increasing mass order.
The interpretation focuses on two dominant particle mixing scenarios: wino-dominated and higgsino-dominated states, which differ in their production cross-sections.
The models assume highly decoupled sleptons, light and third generation squarks and gluinos. The simultaneous production of fully degenerate chargino-1 ($\tilde\chi^{\pm}_{1}$) and/or neutralino-1(2) ($\tilde\chi^{0}_{1,2}$) is considered ($pp\rightarrow \tilde\chi^{\pm}_{1} \tilde\chi^{0}_{1,2}$, $pp\rightarrow \tilde\chi^{\pm}_{1} \tilde\chi^{\pm}_{1}$, $pp\rightarrow \tilde\chi^{0}_{1} \tilde\chi^{0}_{2}$), with an assumed 50\% branching ratio each for decays of $\tilde\chi^{0}_{1,2} \rightarrow \bar{b}t\tau^{-}(b\bar{b}\nu_{\tau})$ and $\tilde\chi^{\pm}_{1} \rightarrow b\bar{b}\tau^{\pm}(\bar{b}t\nu_{\tau})$, as presented in the Figure \ref{fig:SUSYfeynD}.
A previous search~\cite{HDBS-2020-03} for a SUSY model with RPV decays of $\tilde\chi^{\pm}_{1}$ or $\tilde\chi^{0}_{1,2}$ to second- or third-generation of leptons, and third-generation of quarks ($\lambda'_{i33}$) was performed by the ATLAS Collaboration, excluding higgsinos (winos) masses between 200–585 (200–670) GeV.

The paper is structured as follows: Section~\ref{sec:detector} provides an overview of the ATLAS detector, followed by a description of the data and simulated samples in Section ~\ref{sec:samples}. Event reconstruction, object identification, and trigger selections are detailed in Section~\ref{sec:objects}. The search strategy, including signal region definitions and the optimisation of signal classification using neural networks, is outlined in Section~\ref{sec:analysis}. Background estimation and systematic uncertainties are discussed in Sections~\ref{sec:backgrounds} and~\ref{sec:systematics}, respectively. Finally, Section~\ref{sec:result} presents the results along with their statistical interpretation, and Section~\ref{sec:conclusion} concludes the paper.


%

%
\section{ATLAS detector}
\label{sec:detector}

%
\newcommand{\AtlasCoordFootnote}{%
ATLAS uses a right-handed coordinate system with its origin at the nominal interaction point (IP)
in the centre of the detector and the \(z\)-axis along the beam pipe.
The \(x\)-axis points from the IP to the centre of the LHC ring,
and the \(y\)-axis points upwards.
Cylindrical coordinates \((r,\phi)\) are used in the transverse plane,
\(\phi\) being the azimuthal angle around the \(z\)-axis.
The pseudorapidity is defined in terms of the polar angle \(\theta\) as \(\eta = -\ln \tan(\theta/2)\) and is equal to the rapidity $y$ in the relativistic limit.
Angular distance is measured in units of \(\Delta R \equiv \sqrt{(\Delta y)^{2} + (\Delta\phi)^{2}}\).}

The ATLAS detector~\cite{PERF-2007-01} at the LHC covers nearly the entire solid angle around the collision point.\footnote{\AtlasCoordFootnote}
It consists of an inner tracking detector surrounded by a thin superconducting solenoid, electromagnetic and hadron calorimeters,
and a muon spectrometer incorporating three large superconducting toroidal magnets.

The inner-detector system is immersed in a \SI{2}{\tesla} axial magnetic field
and provides charged-particle tracking in the range \(|\eta| < 2.5\).
The high-granularity silicon pixel detector covers the vertex region and typically provides four measurements per track,
the first hit normally being in the insertable B-layer installed before Run~2~\cite{ATLAS-TDR-19,PIX-2018-001}.
It is followed by the silicon microstrip tracker, which usually provides eight measurements per track.
These silicon detectors are complemented by the transition radiation tracker (TRT),
which enables radially extended track reconstruction up to \(|\eta| = 2.0\).
The TRT also provides electron identification information
based on the fraction of hits (typically 30 in total) above a higher energy-deposit threshold corresponding to transition radiation.

The calorimeter system covers the pseudorapidity range \(|\eta| < 4.9\). Within the region \(|\eta|< 3.2\), electromagnetic calorimetry is provided by barrel and
endcap high-granularity lead/liquid-argon (LAr) calorimeters, with an additional thin LAr presampler covering \(|\eta| < 1.8\) to correct for energy loss in material upstream of the calorimeters.
Hadron calorimetry is provided by the steel/scintillator-tile calorimeter, segmented into three barrel structures within \(|\eta| < 1.7\), and two copper/LAr hadron endcap calorimeters. The solid angle coverage is completed with forward copper/LAr and tungsten/LAr calorimeter modules
optimised for electromagnetic and hadronic energy measurements respectively.

The muon spectrometer comprises separate trigger and
high-precision tracking chambers measuring the deflection of muons in a magnetic field generated by the superconducting toroidal magnets.
The field integral of the toroids ranges between \num{2.0} and \SI{6.0}{\tesla\metre}
across most of the detector.
Three layers of precision chambers, each consisting of layers of monitored drift tubes, cover the region \(|\eta| < 2.7\), %
were complemented by cathode-strip chambers in the forward region, where the background is highest.
The muon trigger system covers the range \(|\eta| < 2.4\) with resistive-plate chambers in the barrel, and thin-gap chambers in the endcap regions.

Interesting events are selected by the first-level trigger system implemented in custom hardware,
followed by selections made by algorithms implemented in software in the high-level trigger (HLT)~\cite{TRIG-2016-01}.
The first-level trigger accepts events from the \SI{40}{\MHz} bunch crossings at a rate below \SI{100}{\kHz},
which the high-level trigger further reduces in order to record events to disk at about \SI{1}{\kHz}.

A software suite~\cite{SOFT-2022-02} is used in data simulation, in the reconstruction and analysis of real and simulated data, in detector operations, and in the trigger and data acquisition systems of the experiment.

%


%

%
\section{Data and simulated event samples}
\label{sec:samples}

%
A dataset of $pp$ collisions at a centre-of-mass energy of $\sqrt{s} = \SI{13}{\TeV}$ collected by the ATLAS experiment from 2015 to 2018, and corresponding to integrated luminosity of up to \lumival, is used for this analysis. The number of additional $pp$ interactions per bunch crossing (pile-up) in this dataset ranges from about 8 to 70, with an average of 34. Only events recorded under stable beam conditions and for which all relevant detector subsystems were known to be in a good operating condition are used~\cite{DAPR-2018-01}.

\begin{table}[htpb]
\begin{center}
\caption{\label{tab:mcconfig}
The configurations used for event generation of signal and background processes.
The samples used to estimate the systematic uncertainties are indicated in grey between parentheses.
$V$ refers to the production of an EW boson ($W$ or $Z/\gamma^*$), and $l$ refers to a lepton ($e$, $\mu$ or $\tau$).
The column labelled `Parton shower' refers to the generator used for the parton shower (PS), hadronisation and underlying events. The parton distribution function (PDF) used for the matrix element (ME) is shown.
Tune refers to the underlying-event tune of the PS generator. \textsc{MG5\_aMC}
refers to \textsc{MadGraph5\_aMC@NLO} 2.2 or 2.3~\cite{Alwall:2014hca}; \textsc{Pythia} 8
refers to version 8.2 or 8.3~\cite{Sjostrand:2007gs}; %
\textsc{MePs@Nlo} refers to the method used in \textsc{Sherpa}~\cite{Cascioli:2011va,Gleisberg:2008fv,Schumann:2007mg,Hoeche:2012yf} to match the ME to the PS.
Samples using \textsc{Pythia} 8 have heavy flavour hadron decays modelled by \textsc{EvtGen} 1.2.0~\cite{Lange:2001uf}.
All samples include leading-logarithm photon emission, either modelled by the PS generator or by \textsc{PHOTOS}~\cite{Golonka:2005pn}.
}
\vspace{0.5cm}{\small
\setlength\tabcolsep{1.6pt}
\begin{tabular}{llllll}
\toprule
Process & Generator & ME order & Parton shower & PDF & Tune  \\
\midrule
VLL signal            & \textsc{MG5\_aMC} \cite{Alwall:2014hca}  & LO & \textsc{Pythia} 8\ \cite{Sjostrand:2007gs} & NNPDF3.0 NLO \cite{Ball:2014uwa} & A14 \cite{ATL-PHYS-PUB-2014-021} \\
RPV SUSY signal & \textsc{MG5\_aMC}  & LO & \textsc{Pythia} 8\  & NNPDF3.0 NLO & A14  \\
& & & & \\
$\ttbarj$               & \textsc{PowhegBOX} \cite{Nason:2004rx} & NLO & \textsc{Pythia} 8 & NNPDF3.0 NLO & A14  \\
& {\color{gray}   (\textsc{PowhegBOX})} & {\color{gray}  (NLO)} & {\color{gray} (\textsc{Pythia} 8, $p_{\mathrm{T}}^{\mathrm{hard}}=1$)} & {\color{gray}  (NNPDF3.0 NLO)} &  {\color{gray} (A14)} \\
{\color{gray}}
& {\color{gray}  (\textsc{PowhegBOX})} & {\color{gray}  (NLO)} & {\color{gray}  (\textsc{Herwig7.1.3} \cite{Bellm:2017jjp})} & {\color{gray}  (NNPDF3.0 NLO)} & {\color{gray}  (H7.1 default)} \\
{\color{gray}  ($t\bar{t}b\bar{b}$)}    & {\color{gray} (\textsc{PowhegBOXRES} \cite{Jezo:2018yaf})} & {\color{gray} (NLO)} & {\color{gray} (\textsc{Pythia8})} & {\color{gray} (NNPDF3.0 NLO)} &  {\color{gray} (A14)} \\

& & & & \\
$Z(\to l^+l^-)+\mathrm{jets}$         & \textsc{Sherpa} 2.2.1 \cite{Bothmann:2019yzt}  & NLO & \textsc{Sherpa} & NNPDF3.0 NLO & \textsc{Sherpa} default \\
& & & & \\
$W(\to l\nu)+\mathrm{jets}$               & \textsc{Sherpa} 2.2.1 & NLO & \textsc{Sherpa} & NNPDF3.0 NLO & \textsc{Sherpa} default \\
& & & & \\
$t\bar t t\bar t$      & \textsc{MG5\_aMC} & NLO & \textsc{Pythia} 8 & NNPDF3.1 NLO & A14 \\
& {\color{gray} (\textsc{Sherpa 2.2.11})} & {\color{gray} (NLO)} & {\color{gray} (\textsc{Sherpa})} &  {\color{gray} (NNPDF3.0 NNLO)} & {\color{gray} (\textsc{Sherpa} default)} \\
& & & & \\
$\ttbar W$             & \textsc{Sherpa 2.2.10} & NLO & \textsc{Sherpa} &  NNPDF3.0 NNLO & \textsc{Sherpa} default \\
& {\color{gray} (\textsc{MG5\_aMC})}  & {\color{gray} (NLO)} & {\color{gray} (\textsc{Pythia} 8)} & {\color{gray} (NNPDF3.0 NLO)} & {\color{gray} (A14)}   \\
& & & & \\
\ttH                   & \textsc{PowhegBOX}   & NLO & \textsc{Pythia} 8\ & NNPDF3.0 NLO  & A14 \\
& {\color{gray} (\textsc{PowhegBOX})} & {\color{gray} (NLO)} & {\color{gray} (\textsc{Herwig7.0.4}\cite{Bellm:2015jjp})} &  {\color{gray} (NNPDF3.0 NLO)} &  {\color{gray} (H7 default)} \\
& {\color{gray} (\textsc{MG5\_aMC})}  & {\color{gray} (NLO)} & {\color{gray} (\textsc{Pythia} 8)} & {\color{gray} (NNPDF3.0 NLO)} & {\color{gray} (A14)}   \\
& & & & \\
$\ttbar (Z/\gamma^* \to l^+l^-)$ & \textsc{Sherpa 2.2.11} & NLO & \textsc{Sherpa} &  NNPDF3.0 NNLO & \textsc{Sherpa} default \\ %
& {\color{gray} (\textsc{MG5\_aMC})} & {\color{gray} (NLO)} & {\color{gray} (\textsc{Pythia} 8)} & {\color{gray} (NNPDF3.0 NLO)} & {\color{gray} (A14)}  \\
& & & & \\
Single top:       & \textsc{PowhegBOX} \cite{Re:2010bp,Alioli:2009je} & NLO & \textsc{Pythia} 8 & NNPDF3.0 NLO & A14   \\
$tW$, $s$-, $t$-channel        & & &  \\
& & & & \\
$VH$                   & \textsc{PowhegBOX} & NLO  & \textsc{Pythia} 8 & NNPDF3.0 NLO & A14 \\
$VV$, $qqVV$, $VVV$          & \textsc{Sherpa} 2.2.2  & NLO & \textsc{Sherpa} & NNPDF3.0 NNLO & \textsc{Sherpa} default  \\
$t (Z/\gamma^*)$       & \textsc{MG5\_aMC} & NLO & \textsc{Pythia} 8 & NNPDF2.3 LO  & A14  \\
$tHjb$       & \textsc{MG5\_aMC} & NLO & \textsc{Pythia} 8 & NNPDF2.3 LO  & A14  \\
$t W (Z/\gamma^*)$     & \textsc{MG5\_aMC} & NLO & \textsc{Pythia} 8 & NNPDF2.3 LO  & A14  \\
$t W H$     & \textsc{MG5\_aMC} & NLO & \textsc{Pythia} 8 & NNPDF2.3 LO  & A14  \\
$t\bar t W^+ W^-$      & \textsc{MG5\_aMC} & LO  & \textsc{Pythia} 8 & NNPDF2.3 LO  & A14  \\
$t\bar t t$            & \textsc{MG5\_aMC} & LO & \textsc{Pythia} 8 & NNPDF2.3 LO & A14 \\
$t \bar{t} ZZ$         & \textsc{Madgraph} \cite{Alwall:2014hca} & LO  & \textsc{Pythia} 8 & NNPDF2.3 LO & A14 \\
$t \bar{t} HH$         & \textsc{Madgraph} & LO  & \textsc{Pythia} 8 & NNPDF2.3 LO & A14 \\
$t \bar{t} WH$         & \textsc{Madgraph} & LO  & \textsc{Pythia} 8 & NNPDF2.3 LO & A14 \\
\bottomrule
\end{tabular}
}
\end{center}
\end{table}

Monte Carlo (MC) simulation samples are produced for the different signal and background processes.  Pile-up is modelled by overlaying the simulated hard-scattering event with inelastic $pp$ events generated with \PYTHIA[8.186]~\cite{Sjostrand:2007gs} using the \NNPDF[2.3lo] set of parton distribution functions (PDF)~\cite{Ball:2012cx} and the A3 set of tuned parameters~\cite{ATL-PHYS-PUB-2016-017} (referred to as the `tune').  The MC samples are produced using either a detailed ATLAS detector simulation~\cite{SOFT-2010-01} based on \textsc{Geant4}~\cite{Agostinelli:2002hh} or a faster simulation where the full \textsc{Geant4} simulation of the calorimeter response is replaced by a parameterisation of the shower shapes~\cite{SOFT-2010-01}. In both cases, the simulated events are processed through the same reconstruction software as the dataset of $pp$ collisions.  For all samples of simulated events, except those generated using the \SHERPA generator~\cite{Cascioli:2011va,Gleisberg:2008fv,Schumann:2007mg,Hoeche:2012yf}, the \EVTGEN%
program~\cite{Lange:2001uf} is used to describe the decays of bottom and charm hadrons.
The top-quark mass $m_{t}$ is set to $172.5$~GeV and the Higgs boson mass $m_{H}$ is set to $125$~GeV in all relevant samples.

Corrections are applied to the simulated events so that the particle candidates' selection efficiencies, energy scales and energy resolutions match those determined from data control samples.  The simulation samples are normalised to their cross-sections, and generated to the highest order available in perturbation theory.

Signal samples for the \vll model~\cite{Cornella:2021sby,DiLuzio:2018zxy,Baker:2019sli} are generated for VLL masses in the range $200$--$1500${~\GeV} in intervals of $\SI{100}{\GeV}$.
The production via \Zp is suppressed by setting the \Zp mass to 100 TeV and its couplings to 0. Similarly, $G'$ is not relevant in this search, as this gauge boson couples only to coloured particles. Its mass is therefore set to 5~TeV, in order to suppress processes via its exchange. To suppress the EW production of VLQs, all VLQ masses are set to 3.5~TeV. To accommodate flavour anomalies, the model favours vector leptoquark \Umu and VLL decays to third-generation fermions, as shown in Ref. \cite{Cornella:2021sby}. Decays to the first two generations are suppressed by fixing the corresponding couplings to~0, while \Umu couplings to third-generation fermions are set to~1. The \Umu width is constrained to 500 GeV by adjusting its couplings to VLLs and VLQs to be between -0.5 and~1. The \Umu couplings to the SM fermions and mass are chosen to also align with flavour anomalies, resulting in two scenarios \cite{Cornella:2021sby}: one not allowing right-handed currents ($\beta^{b\tau}_R$ = 0) and one with maximal right-handed currents ($\beta^{b\tau}_R$ = -1).
Due to similar kinematics and cross-section for the two scenarios, only the $\beta^{b\tau}_R$ = 0 scenario is investigated.
The mass and universal coupling of \Umu are set to 3.5 TeV and 3, respectively. This brings to off-shell \Umu decays, allows for consistency with flavour anomalies, and preserves compatibility with the model previously studied by the CMS Collaboration \cite{CMS-B2G-21-004}. The VLL widths are in the very narrow range of $10^{-2}$--$10^{-6}$ \GeV, resulting in 3-body VLL decays.
VLL pair production and decay was done using {\scshape MadGraph5\_aMC@NLO 2.2.3} \cite{Alwall:2014hca,Frederix:2012ps} at leading order (LO) with the {\scshape Pythia 8.186}~\cite{Sjostrand:2007gs}%
interface for the parton shower (PS), hadronisation and underlying event modelling, utilising the A14 tune~\cite{ATL-PHYS-PUB-2014-021}.
The matrix element (ME) is calculated at tree level, with emission of up to two additional partons. The PDF set used is \NNPDF{3.0}{\scshape lo}~\cite{Ball:2014uwa}, and the ME-PS matching uses the {\textsc CKKW-L} prescription~\cite{Lonnblad:2001iq,Lonnblad:2011xx}.
The VLL pair production cross-section is calculated using {\scshape MadGraph} at next-to-leading order (NLO) accuracy in Quantum Chromodynamics (QCD).
The RPV SUSY signal samples are generated using {\scshape MadGraph5\_aMC@NLO~2.9.9} at LO, with up to two initial state radiation jets at LO in QCD, and the {\scshape Pythia 8.307}~\cite{Bierlich:2022pfr} interface for the PS, hadronisation and underlying event modelling, utilising the A14 tune~\cite{ATL-PHYS-PUB-2014-021}. The matching scale is set to 1/4 of the produced SUSY particle mass.
The cross-section is calculated at NLO in QCD, with added resummation of soft gluon emission at next-to-leading-logarithmic accuracy (NLL) using Resummino~\cite{Fuks:2013vua}. The nominal cross-section and uncertainty are calculated from an envelope of calculations using different PDF sets and factorisation and renormalisation scales following the prescription described in Ref.~\cite{Butterworth:2015oua}.

Samples for \ttbar production, single top production, vector boson ($V$) production in association with the Higgs boson ($VH$), and the associated production of a top-quark pair and a Higgs boson ($\tth$) are generated using the NLO generator \textsc{Powheg-Box}~\cite{Frixione:2007nw,Nason:2004rx,Frixione:2007vw,Alioli:2010xd,Alioli:2009je} %
with the \NNPDF{3.0}{\scshape nlo} PDF set; interfaced with {\scshape Pythia 8.2} with the A14 tune. %
The $h_\text{damp}$ parameter, which controls the \pt of the first additional emission beyond the Born configuration and therefore regulates the high-$\pt$ radiation, is set to $3(2m_t + m_{H})/4$ in the \ttH sample and to $1.5 m_t$ in the \ttbarj and single top samples. %

The $Z(\to l^+l^-)+\mathrm{jets}$ and $W(\to l\nu)+\mathrm{jets}$ events are simulated using the \sherpa~2.2.1 generator~\cite{Bothmann:2019yzt} with NLO-accurate MEs for up to two partons and LO-accurate MEs for up to four partons. The MC yields are normalised to the next-to-NLO (NNLO) cross-sections, computed using \textsc{Fewz}~\cite{Anastasiou:2003ds} with the {\scshape MSTW2008NNLO}~\cite{Martin:2009iq} PDF set.

The sample used to model the \ttW (\ttdy) background process is generated using \Sherpa~2.2.10 (\Sherpa~2.2.11), where the ME is calculated for up to one (zero) additional parton at NLO in QCD and up to two partons at LO in QCD using \textsc{Comix}~\cite{Gleisberg:2008fv} and \textsc{OpenLoops}~\cite{Cascioli:2011va} and merged with the \textsc{Sherpa} PS~\cite{Schumann:2007mg} using the \textsc{MePs@Nlo} prescription~\cite{Hoeche:2012yf}.  A {\scshape CKKW} merging scale of $30$~\gev is used for the \ttW sample. These samples are generated using the {\scshape NNPDF3.0nnlo} %
PDF set. Both the factorisation and renormalisation scales are set to $\mu_R = \mu_F =H_\textrm{T}$/2 in the \ttW sample, where $H_\textrm{T}$ is defined as the scalar sum of the transverse masses $\sqrt{p_\textrm{T}^2+m^2}$ of all final state particles. The \ttW LO EW contributions are obtained from a dedicated sample simulated with \Sherpa~2.2.10 and are added together with the NLO QCD sample described above. %

Diboson ($WZ$, $ZZ$ and $WW$ including off-shell productions) background processes are simulated with \textsc{Sherpa}~2.2.2.  The ME is calculated using $\textsc{Comix}$ and \textsc{OpenLoops} at NLO accuracy in QCD for up to one additional parton and at LO accuracy for up to three additional partons, and merged with the \textsc{Sherpa} PS model using the \textsc{MePs@Nlo} prescription. The {\scshape NNPDF3.0nnlo} set of PDFs is used, along with the dedicated set of tuned PS parameters developed by the \SHERPA authors. %

Table~\ref{tab:mcconfig} shows the configurations for all MC samples used in this analysis, with the samples in parentheses indicating those used to estimate the systematic uncertainties. %
%


%

%
\section{Event reconstruction and object identification}
\label{sec:objects}

%
%
Events are selected using requirements involving the reconstructed primary interaction vertex, triggers, and reconstructed jets, \bjets and \tauh candidates.  Additionally, events with light leptons are used to estimate the background from $t\bar{t}$ and \Zjets processes.
The primary interaction vertices of the $pp$ collision events are reconstructed based on at least two tracks that are consistent with originating from the beam collision region and have \pt higher than $500$~MeV. Of all primary vertex candidates, the one associated with the highest sum of squared $\pt$ for its associated tracks is taken as the hard-scatter primary vertex~\cite{ATL-PHYS-PUB-2015-026}.
Events are required to satisfy data quality criteria to remove significant noise in the calorimeters or data corruption~\cite{DAPR-2018-01}. Additional quality cuts are applied to discard jet candidates originating from non-collision sources or detector noise, as well as to remove muon candidates with significantly worse momentum resolution.

Electron candidates are reconstructed from energy clusters in the electromagnetic calorimeter that are associated with charged particle tracks reconstructed in the inner detector~\cite{EGAM-2018-01,EGAM-2021-01,EGAM-2021-02}. These candidates are required to satisfy $\pt > 10$ GeV and $|\eta_\textrm{cluster}| < 2.47$, and they must not be in the transition region between different electromagnetic calorimeter components, $1.37 < |\eta_\textrm{cluster}| < 1.52$. A multivariate likelihood discriminant combining shower shape and track information is used to distinguish real electrons from hadronic showers~\cite{EGAM-2018-01,EGAM-2021-02}. A loose electron discriminant working point, denoted as `LooseLH', is employed for the preselection, which is further tightened to `TightLH' working point in certain analysis regions. %

Muon candidates are reconstructed by combining inner-detector tracks with track segments or full tracks in the muon spectrometer \cite{MUON-2022-01,MUON-2018-03}.  %
Candidates are preselected with the Loose identification working point, and they are required to satisfy $\pt > 10$~GeV and $|\eta| < 2.5$. %

The electron (muon) candidate track is associated with the primary vertex by restricting the significance of its transverse impact parameter, $d_0$, to be such that $|d_0/\sigma(d_0)|<5 (3)$, where $\sigma(d_0)$ denotes the measured uncertainty in $d_0$. Furthermore, the longitudinal impact parameter $z_0$ must satisfy $|z_0 \sin\theta|<0.5$~mm, where $\theta$ is the polar angle of the track.  The electron or muon candidates are also required to be isolated in the tracker and in the calorimeter. A track-based lepton isolation criterion is defined by calculating the quantity $I_R = \sum \pt^{\textrm{trk}}$, where the scalar sum includes all tracks (excluding the lepton candidate itself) within a cone defined by $\Delta R<R_{\textrm{cut}}$ around the direction of the lepton.  The value of $R_{\textrm{cut}}$ is the smaller of $r_{\textrm{min}}$ and $10~\mathrm{\GeV}/\pt^\ell$, where $r_{\textrm{min}}$ is set to 0.2 (0.3) for electron (muon) candidates and where $\pt^\ell$ is the lepton $\pt$. All lepton candidates must satisfy $I_R/\pt^\ell < 0.15$.
Additionally, electrons (muons) are required to satisfy a calorimeter-based isolation criterion: the sum of the transverse energy within a cone of size $\Delta R=0.2$ around the lepton, after subtracting the contributions from pile-up and the energy deposit of the lepton itself, is required to be less than 20\% (30\%) of $\pt^\ell$.

The \tauh candidates are reconstructed from energy clusters in the calorimeters and associated inner-detector tracks~\cite{ATL-PHYS-PUB-2022-044}. %
They are required to have either one or three associated tracks (referred to as 1-prong and 3-prong $\tauh$ candidates, respectively), with a total charge of $\pm 1e$.
The candidates are required to satisfy $\pt > \SI{20}{\GeV}$ and $|\eta|<2.5$, excluding the electromagnetic calorimeter's transition region.
A recurrent neural network discriminant using calorimeter- and tracking-based variables is used to identify real $\tauh$ candidates and suppress mis-reconstructed jet backgrounds~\cite{ATL-PHYS-PUB-2019-033}. Three identification working points of 1-prong and 3-prong \tauh candidates are used for different levels of efficiencies and background rejections.  The selected $\tauh$ candidates are referred to as `Loose', `Medium', and `Tight', respectively~\cite{ATL-PHYS-PUB-2019-033}.
The loose working point has a target efficiency of $85\%$ ($75\%$) for 1-prong (3-prong) \tauh candidates, with an expected rejection factor against light-jets of 21 (90). The corresponding efficiencies and rejections for the medium working point are $75\%$ ($60\%$) and 35 (240) for 1-prong (3-prong) \tauh candidates, respectively. The tight working point has average efficiency of $60\%$ ($45\%$) for 1-prong (3-prong) \tauh candidates, respectively.
Electrons that are reconstructed as 1-prong $\tauh$ candidates are removed using a boosted decision tree algorithm~\cite{ATL-PHYS-PUB-2019-033}. %
The $\tauh$ reconstruction and identification efficiencies and the $\tauh$ energy scale in the simulation are calibrated to those measured in a data control sample of $Z \to \tau^+\tau^-$ events~\cite{ATLAS-CONF-2017-029}, and the associated uncertainties are considered in the analysis.

The constituents for jet reconstruction are identified by combining measurements from both the inner-detector and the calorimeter using a particle flow (PFlow) algorithm~\cite{PERF-2015-09,JETM-2018-05}. Jet candidates are reconstructed from these PFlow objects using the anti-$k_t$ algorithm~\cite{Cacciari:2008gp,Cacciari:2005hq} with a radius parameter of $R=0.4$. They are calibrated using simulation with corrections obtained from \textit{in situ} techniques in data~\cite{JETM-2018-05}. Only jet candidates with a  $\pT> 25$~\GeV~and within $|\eta|<2.5$ are selected. In order to reduce the effect from pile-up, each jet with $\pT<60$~\GeV~and $|\eta|<2.4$ is required to satisfy the tight working point of the jet vertex tagger (JVT)~\cite{ATLAS-CONF-2014-018} criteria used to identify the jets as originating from the selected primary vertex. %

Jets containing $b$-hadrons are identified ($b$-tagged) via the \texttt{DL1r} multivariate discriminant combining information from the impact parameters of displaced tracks with topological properties of secondary and tertiary decay vertices reconstructed within the jet~\cite{FTAG-2018-01,FTAG-2019-07}. Each jet can be sorted into a pseudo-continuous binned distribution of the DL1r discriminant score, where each bin corresponds to a certain range of $b$-tagging efficiencies defined by the working points corresponding to an average efficiency of 60\%, 70\%, 77\% and 85\% for $b$-jets with $\pt > 20$~GeV and $|\eta| < 2.5$ in $t\bar t$ events.
The $85\%$ working point is used in the event selection of channels used for background estimation in this analysis ($\geq\!1\ell$ channels). The $77\%$ working point is applied in channels of both signal and control regions ($0\ell$ channels). Operating at the $77\%$ working point, the tagging algorithm achieves a rejection factor of about 170 against light-jets (jets that do not contain $b$- or $c$-hadrons) and of approximately 5 against $c$-jets. The pseudo-continuous $b$-tagging information is used as an input variable in the signal region optimisation.

The \MET is defined as the magnitude of the negative vector sum of the transverse momenta of all objects in the event. It also includes a term for the transverse momentum coming from all soft particles that are not associated with any of the objects in the event \cite{JETM-2020-03}, which is calculated from inner detector tracks that can be matched to the primary vertex.
In order to resolve ambiguities in object reconstruction, an algorithm is used to avoid double counting of two different close-by objects that pass the selection requirements and overlap with each other. The overlapping objects are removed according to the following procedure. Electrons and muons that pass the loose selections are considered for overlap removal, together with calorimeter jets which pass the JVT requirement and \tauh candidates identified using the Loose criteria. %
If the electron and muon share a track, the electron is removed if the muon is associated with a signature in the muon spectrometer, otherwise the muon is removed. For a surviving electron which overlaps with a jet within  $\Delta R <$ 0.2, the jet is removed. If an electron and a jet overlap within $\Delta R <$ 0.4, the electron is removed. If a muon and a jet overlap with $\Delta R <$ 0.2, the jet is removed, unless the number of tracks in the jet is more than 2. If a muon and a jet overlap with $\Delta R <$ 0.4, the muon is removed. If a surviving electron overlaps with a loose \tauh within $\Delta R < 0.2$, the electron is kept and the \tauh object is removed. If a surviving muon overlaps with a loose \tauh within $\Delta R < 0.2$, the muon is kept and the \tauh object is removed. If a surviving jet overlaps with a loose \tauh within $\Delta R < 0.2$, the jet is removed.

\subsection{Trigger selection}
The analysis employs different trigger strategies for the $0\ell$ events, which contain at least one \tauh and no isolated light leptons.  The $\geq\!1\ell$ channels, which include at least one isolated light lepton and are used only for the background estimations, are selected using single-light-lepton (SLT) or double-light-lepton (DLT) triggers.

The various types of triggers considered in the analysis are described in the following. At the HLT, the \MET triggers (MET)~\cite{TRIG-2019-01} have minimum thresholds ranging from 90~\GeV to 110~\GeV depending on the data-taking period. The single-\tauh triggers (STT)~\cite{ATLAS-CONF-2017-061} select a \tauh candidate satisfying HLT quality criteria and with a \pt threshold ranging from $80$~\GeV to $160$~\GeV, where the $160$~\GeV threshold is used for the majority of the data-taking period.  The di-\tauh (DTT) triggers have kinematic thresholds on two \tauh candidates, as well as on the leading~\footnote{In this paper, the offline physics objects are ordered in \pt, with the leading object referring to the one with the highest \pt.} and sub-leading jet \pt in certain triggers depending on the data-taking period, as described in Ref.~\cite{HDBS-2018-40}.  The \pt threshold of the leading (sub-leading) \tauh is $35 (25)$~\GeV across all data-taking periods. For the years 2017--2018, a subset of triggers introduces an additional requirement on the angular separation $\Delta R(\tauh,\tauh)$ between two \tauh candidates. This subset also includes a single jet requirement. These requirements were designed to reduce trigger rates by suppressing back-to-back \tauh pairs. There are \bjet triggers (BJET)~\cite{TRIG-2018-08} employed during the data taking period from 2016--2018, which require at least one jet with \pt above $100$~\GeV or $110$~\GeV depending on the data-taking period and two $b$-tagged jets identified with the HLT algorithm based on multivariate analysis techniques.  The $b$-tagged jets have \pt above the $35$--$55$~\GeV range depending on the data-taking year.
Furthermore, the data are selected using a combination of SLT or DLT triggers, with requirements on the identification, isolation, and \pt of the light leptons to maintain a high efficiency across the full momentum range while controlling the trigger rates~\cite{TRIG-2018-01,TRIG-2018-05}. The electron (muon) SLT has the lowest \pt thresholds of 24--26~(20--26)~\GeV, while the electron (muon) DLT has the lowest \pt threshold requirements of 12--24~(14--22)~\GeV for the leading lepton, depending on the data-taking period and trigger combination. %

\begin{figure}[tbh!]
\centering
\includegraphics[width=1.\textwidth]{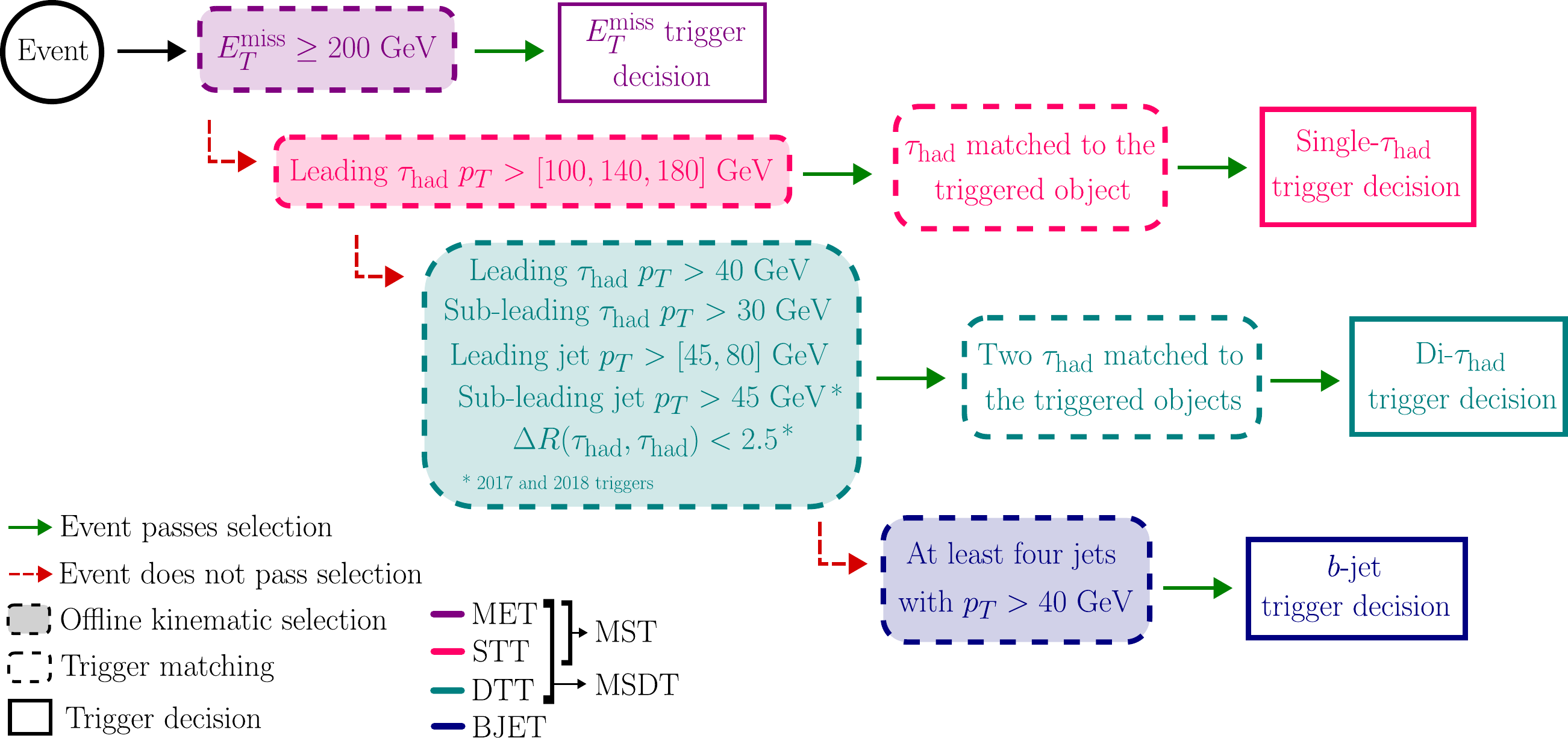}
\caption{Trigger selection strategy for $0\ell$ channels using multiple triggers to maximise signal acceptance, where the kinematic selections in the analysis are chosen in the plateau region of the trigger efficiency, followed by corresponding trigger matching and trigger decisions.
The \MET (MET) trigger category targets events with \MET$\geq\!200$ \GeV, the single-\tauh (STT) triggers select events with single \tauh of \pt above the thresholds indicated in the diagram, the di-\tauh (DTT) triggers recover events with two \tauh with relatively lower thresholds, and the \bjet (BJET) triggers recover events that have low \tauh \pt but a large multiplicity of \bjets. The MST category refers to events that pass either the STT or the MET trigger chain. Similarly, events that belong to the STT, the DTT, or the MET trigger chains are collectively labelled as MSDT.  Variations in the offline thresholds are due to changes in trigger thresholds across different data-taking periods.}
\label{fig:trigger}
\end{figure}

In the $0\ell$ channels, events are selected using a combination of triggers in order to enhance the signal acceptance. These triggers are used with a set of requirements on the kinematic thresholds of the \tauh, light-jet, or \bjet candidates, or the \MET. Events are classified into orthogonal categories based on the kinematic selections which are chosen in the plateau region of trigger efficiency. This methodology was previously used in Refs.~\cite{SUSY-2020-16,HDBS-2018-40}. Furthermore, in the offline analysis, the reconstructed \tauh are geometrically matched to the HLT \tauh that triggered the event. A diagram laying out the trigger strategy in the $0\ell$ channels can be found in Figure ~\ref{fig:trigger}.

There are four different trigger chains based on the type of the trigger used in the $0\ell$ event selection. The selection starts with MET triggers, which are fully efficient for events with \MET~$> 200$~\GeV.  No efficiency corrections are applied to simulated events selected with MET triggers. For events with \MET~$< 200$~\GeV, the STT triggers are used, where the reconstructed \tauh candidate satisfy \pt thresholds, as illustrated in Figure~\ref{fig:trigger}, with efficiencies at plateau.
The selection threshold is up to $20$~\GeV above the trigger threshold.  Signal events with one \tauh or two or more \tauh that pass the orthogonal offline selections have an STT trigger efficiency ranging from $88\%$ to $92\%$, depending on the \tauh multiplicity and signal mass.  The events that do not meet the criteria for MET and STT triggers are recovered with the use of the DTT triggers that have requirements on the \pt of two \tauh candidates and the leading jet.  The reconstructed \tauh candidate \pt must be $5$~\GeV above the trigger threshold applied to the \tauh at HLT, as highlighted in Figure~\ref{fig:trigger}, while the leading reconstructed jet must have $\pt > 80$~\GeV. For the events selected with DTT triggers that required two or more jets, a lower threshold of $45$~\GeV is used for the leading jet. The DTT trigger efficiency for signal events with two \tauh passing the offline selection ranges from $82\%$--$90\%$, depending on the signal mass.  The STT and DTT trigger efficiencies are corrected in the simulated events.  Other remaining events that fail to satisfy the above triggers are selected with BJET triggers if the events contain at least four jets with $\pt > 40$~\GeV. The BJET trigger efficiency ranges from $56\%$ to $90\%$, depending on the signal mass. Events from the BJET trigger category have efficiency corrections related to $b$-tagging.  These corrections are evaluated using a technique previously applied in Ref.~\cite{SUSY-2020-16}.
SLT and DLT triggers are used to select events with two or more light leptons that have \MET $<200$~\GeV.  The reconstructed leptons in the event are required to be matched to the triggered objects at the HLT from the SLT or DLT, with reconstructed lepton \pT 1~\GeV higher than the threshold set on the HLT trigger object. To allow full trigger efficiency, the events having an \MET of at least $200$~\GeV are selected with the MET trigger. These events are used in the determination of kinematic dependent corrections for the $t\bar{t}$ and \Zjets processes, and for the estimations of the scale factors for the misidentified \tauh, as will be discussed in Section~\ref{sec:backgrounds}.

%
%
%
%


%

%
\section{Search strategy}
\label{sec:analysis}

\begin{figure}[htbp]
\includegraphics[width=\textwidth]{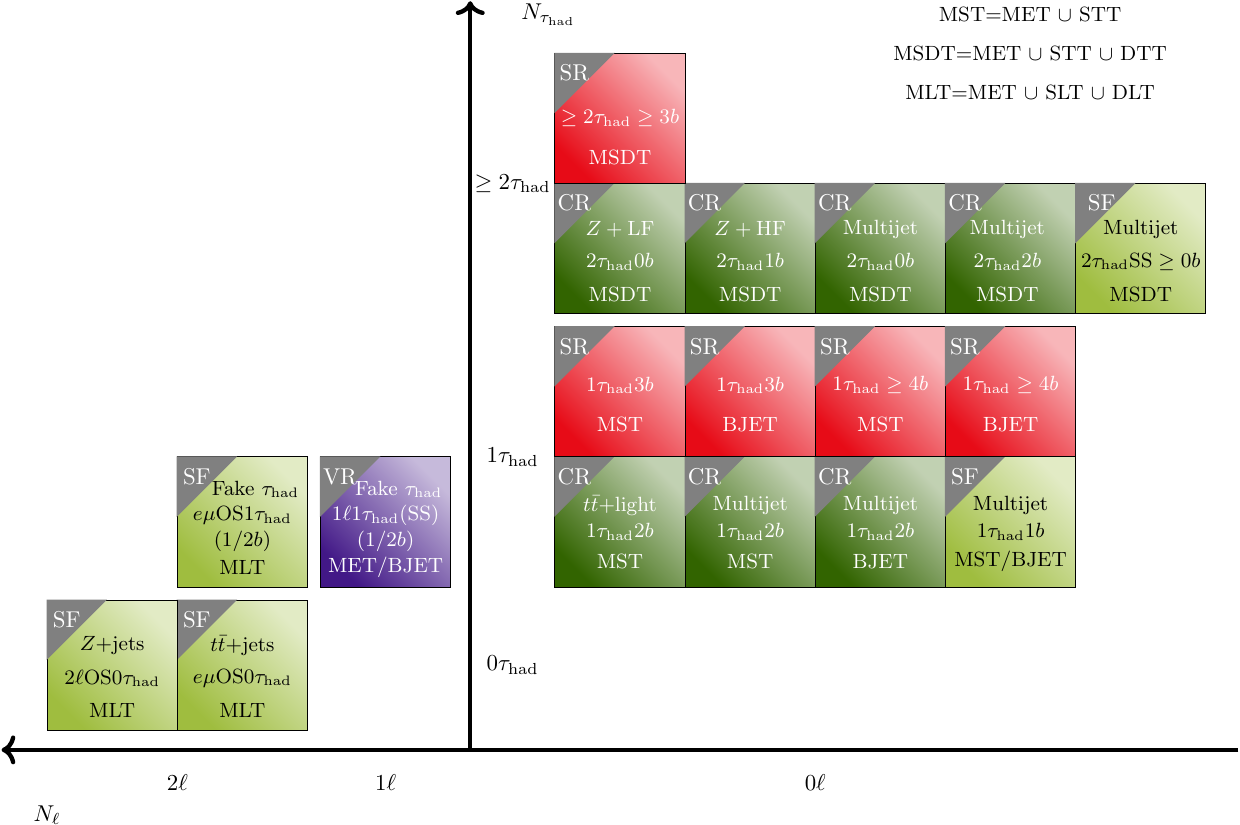}
\caption{A sketch of the analysis regions considered in this \vll VLL search. The regions labelled with `SR' (`CR') refer to the signal (control) regions, as summarised in Table~\ref{tab:tauonly_srs} (\ref{tab:taucrs}).  They are used in the simultaneous profile likelihood fit, as discussed in Section~\ref{sec:result}. The regions labelled with `SF' are used for the derivation of correction factors which depend on event kinematics for the \ttbarj, \Zjets, fake \tauh and multijet background. The validation region `VR' is used to verify the modelling of the fake \tauh background mainly originating from \ttbarj and \WZjets processes. The selections for `SF' and `VR' regions are summarised in Table~\ref{tab:ttbarZjetsFakeTauCRs}. The \ZLF (\ZHF) region is enriched in $Z$ boson events without (with) additional $b$-tagged jets.
A few analysis regions with different trigger selections are labelled with abbreviated notations: MST for the combined MET and STT trigger categories, MSDT for the combined MET, STT and DTT trigger categories, and MLT for the combination of MET triggers with SLT or DLT triggers.}
\label{fig:AR_schematic}
\end{figure}

Based on the considered model, the analysis targets VLL pair production in the final state with one or more \tauh candidates and three or more \bjets.  A set of orthogonal analysis regions are defined to constrain the background and to maximise the signal sensitivity. Given the different thresholds and topologies of the triggers considered, the analysis regions are categorised with a combination of trigger selections to capture events with a single high-\pt \tauh candidate, multiple \tauh candidates with softer \pt, large multiplicity of \bjets, or high \met.  Furthermore, the signal events are split depending on the number of \tauh and the number of \bjets, leading to five signal regions (SRs) in total, as highlighted in the schematic diagram of Figure~\ref{fig:AR_schematic} and described in Section~\ref{subsec:SRs}. These regions are devoid of any isolated light leptons satisfying loose identification with a minimum \pt of 10 GeV. Each of these categories shows a different composition of background, with main contributions from \ttbarj (with real \tauh or with jets misidentified as \tauh, also called fake \tauh), \Zjets and multijet productions. Multivariate analysis techniques are used to search for a VLL pair in all signal regions, and the distribution of a neural network (NN) discriminant score is used in the final likelihood fit discussed in Section~\ref{sec:result} to extract the signal yield. %

The data-driven estimation of \ttbarj and \Zjets backgrounds is carried out in two main steps.  First, kinematic-dependent corrections are derived to address MC mismodelling using dedicated scale-factor (SF) regions and are then applied to the respective background events in all relevant analysis regions. Dedicated regions enriched in either \ttbarj or \Zjets featuring two light leptons, as shown in Figure~\ref{fig:AR_schematic}, are used only for deriving these corrections. In a second stage, the normalisation of individual processes, categorised into light- and heavy-flavour additional jets, is determined from the likelihood fit. The associated production of \ttbar with additional \bjets or \cjets from QCD emissions is known to be poorly modelled in the MC simulations~\cite{CMS-TOP-22-009,TOPQ-2017-01}. Three  individual categories are hence considered for the \ttbarj process, and the corresponding normalisations are determined from the fit to data, as discussed in Section~\ref{sec:ttbar}.  %

Dedicated control regions, as shown in Figure~\ref{fig:AR_schematic}, are defined to estimate the multijet background separately in events with $1\tauh$ and $\geq\!2\tauh$ using data-driven techniques as described in Section~\ref{sec:backgrounds}.

\subsection{Signal region definitions}
\label{subsec:SRs}
\begin{figure*}[t]
\centering
\includegraphics[width=0.55\textwidth]{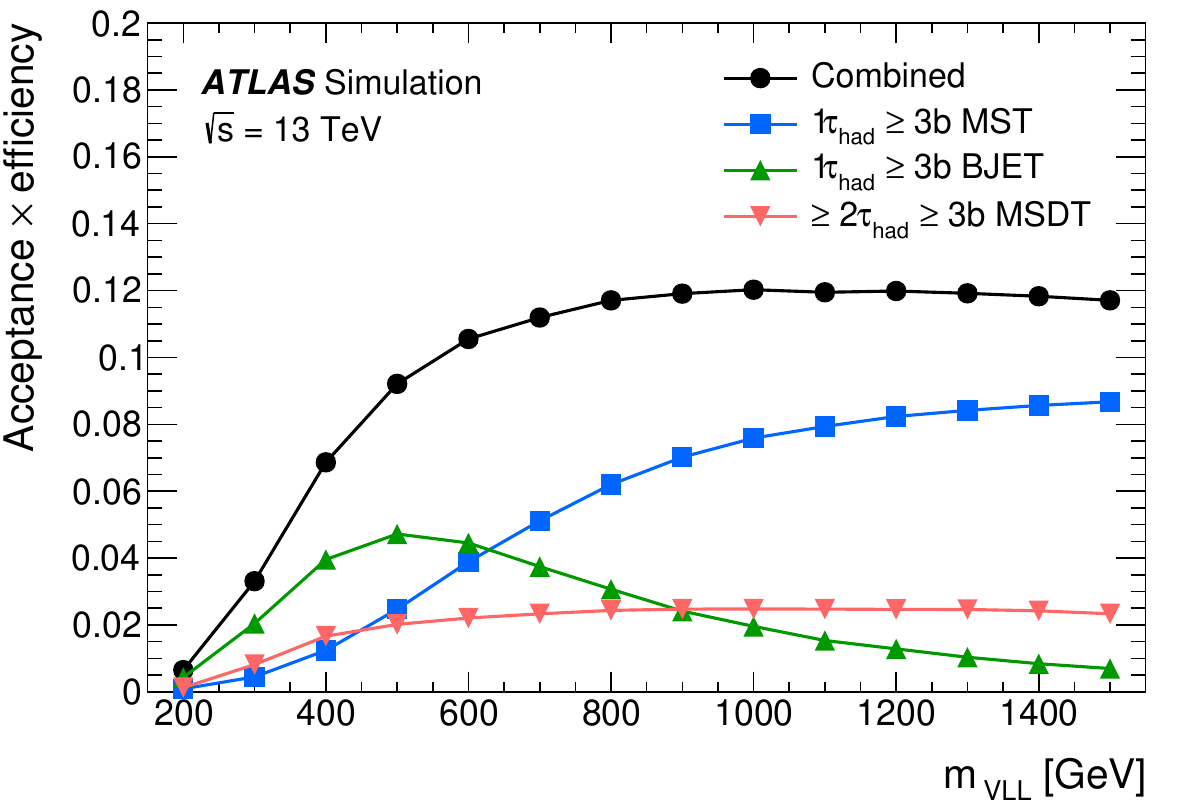}
\caption{The expected acceptance times efficiency (including object identification and reconstruction, trigger selection, and event selection) for VLL signal as a function of \mVLL for the combined signal regions as well as for three signal region categories: $1\tauh\geq\!3b$ MST, $1\tauh\geq\!3b$ BJET and $\geq\!2\tauh\geq\!3b$ MSDT.  The region $1\tauh\geq\!3b$ MST refers to the combination of the $1\tauh 3b$ MST and $1\tauh \geq\!4b$ MST signal regions, while the region $1\tauh\geq\!3b$ BJET refers to the combination of the $1\tauh 3b$ BJET and $1\tauh\geq\!4b$ BJET signal regions.
}
\label{fig:signalAccEff}
\end{figure*}

The signal events selected with at least one \tauh satisfying loose identification with \pt above $\SI{20}{\GeV}$ and three or more \bjets with \pt$\SI{>25}{\GeV}$ are required to pass either of the four trigger categories: STT, DTT, MET and BJET. The \tauh identification criteria are further tightened depending on the \tauh multiplicity and the trigger selections, as summarised in Table~\ref{tab:tauonly_srs} for the five signal regions.  Events consisting of a single \tauh candidate passing the high \pt threshold STT or MET trigger are dominated by backgrounds from \ttbarj with real or fake \tauh, while those passing the BJET trigger category have substantial contributions from multijet events. Due to these differences in the background composition, the single \tauh events are categorised into MST and BJET triggers, which are further split into two regions: one selecting exactly three $b$-tagged jets ($3b$), and the other at least four $b$-tagged jets ($\geq\!4b$) thus allowing an improved estimation of the background with higher heavy-flavour jet content. The MST refers to the combined MET and STT trigger category.  No further splitting is done in the region with two or more \tauh, because of limited statistics.  This region is referred to as $\geq\!2\tau\geq\!3b$ MSDT, where the acronym MSDT refers to the combined MET, STT and DTT trigger categories.  The events selected by the BJET triggers are not considered in this region given their negligible additional contribution to the signal acceptance.  The main background source in this region is \ttbarj with substantial contributions from processes with fake \tauh candidates. %
The signal acceptance times efficiency is presented in Figure~\ref{fig:signalAccEff} for the three categories of selections: $1\tauh\geq\!3b$ MST, $1\tauh\geq\!3b$ BJET, and $\geq\!2\tauh\geq\!3b$ MSDT, illustrating complementary contributions to the signal acceptance from various trigger combinations.  The $1\tauh\geq\!3b$ MST region is the combination of the $1\tauh 3b$ MST and $1\tauh \geq\!4b$ MST signal regions, whereas the region $1\tauh\geq\!3b$ BJET is the combination of the $1\tauh 3b$ BJET and $1\tauh\geq\!4b$ BJET signal regions.

\begin{table}[b]
\caption{Summary of the selections in each of the five signal regions.
The $\mathrm{`OS'}$ refers to an opposite-charge \tauh pair. A few combinations of trigger selections used in the analysis regions are labelled with abbreviated notations: MST for the combined MET and STT trigger categories, and MSDT for the combined MET, STT and DTT trigger categories. The last row refers to the neural network (NN) score distributions in the specified event categories considered for the optimisation of the signal classification, as described in Section~\ref{subsec:NNs}.}
\begin{center}
\resizebox{1\linewidth}{!}{%
\begin{tabular}{c|c|c|c|c|c|c|c}\toprule
Signal Region & 1$\tauh 3b$ MST & 1$\tauh \geq\!4b$ MST & 1$\tauh 3b$ BJET & 1$\tauh \geq\!4b$ BJET & \multicolumn{3}{c}{$\geq\!2\tauh\geq\!3b$ MSDT} \\ \midrule
Triggers & \multicolumn{2}{c|}{MET $\cup$ STT} & \multicolumn{2}{c|}{BJET} & MET & STT & DTT \\
Number of \tauh & \multicolumn{4}{c|}{1} & $\geq\!2$ & \multicolumn{2}{c}{$2$ (OS) or $\geq\!3$} \\
\tauh identification & \multicolumn{4}{c|}{Tight} & $\geq\!1$ Loose & $\geq\!1$ Medium & $\geq\!2$ Medium \\
Number of $b$-tags $@77\%$ & $3$ & $\geq\!4$ & $3$ & $\geq\!4$ & \multicolumn{3}{c}{$\geq\!3$} \\
Number of jets & \multicolumn{4}{c|}{$\geq\!4$} & \multicolumn{3}{c}{$\geq\!3$} \\
NN score distribution & \multicolumn{2}{c|}{NN ($1\tau_{\mathrm{had}}$ MST)} & \multicolumn{2}{c|}{NN ($1\tau_{\mathrm{had}}$ BJET)} & \multicolumn{3}{c}{NN ($\geq\!2\tau_{\mathrm{had}}$ MSDT)} \\ \bottomrule
\end{tabular}
}
\label{tab:tauonly_srs}
\end{center}
\end{table}

\subsection{Optimisation of signal classification with neural networks}
\label{subsec:NNs}
Three NNs are optimised based on the type and composition of backgrounds in specific signal regions, while the statistical interpretation is performed using the score distribution in the five SRs as the final discriminant variables to extract the signal. The NNs are parameterised in terms of the VLL mass to obtain optimal signal sensitivity and smooth interpolations across the mass range~\cite{Baldi:2016fzo}. This approach is equivalent to training the signal against the background for each signal mass hypothesis separately.  Each NN is conditioned on a signal mass parameter which scans the range of VLL masses under investigation.  For each VLL mass point considered, a distinct NN score distribution for the signal, the SM background, and data is evaluated. Hence, the statistical analysis is performed separately for each assumed mass value of the VLL signal model, as the data and background distributions of the NN score change with the mass value.  The NN consists of two hidden dense layers with 35 nodes each and an output layer with two nodes, implemented in Keras~\cite{chollet2015keras} with the Tensorflow ~\cite{tensorflow} backend.

\begin{table}[t]
\caption{Summary of variables used as input to the parameterised neural network trained for three different event categories.}
\begin{center}
\begin{tabular}{cccc}\toprule
\multicolumn{1}{c}{Variable} & \multicolumn{3}{c}{Event category} \\ \cline{2-4}
& 1$\tauh \geq\!3b$ MST & 1$\tauh \geq\!3b$ BJET & $\geq\!2\tauh \geq\!3b$ MSDT \\ \midrule
\ntrigbuck & \checkmark &  & \checkmark \\
\tauptlead   & \checkmark & \checkmark & \checkmark \\
\tauptsublead   &  &  & \checkmark \\
\tauptthirdlead   &  &  & \checkmark \\
\njets & \checkmark & \checkmark & \checkmark \\
\ntaus &  & & \checkmark \\
\HTjets & \checkmark & \checkmark & \checkmark \\
\met & \checkmark & \checkmark & \checkmark \\
\mtautau &  &  & \checkmark \\
\mtaub & \checkmark & \checkmark & \checkmark \\
\mbbtau & \checkmark & \checkmark & \checkmark \\
\mbbmet & \checkmark & \checkmark & \checkmark \\
\sumPCB & \checkmark &  & \checkmark \\
\sumPCT & & & \checkmark \\
\minDphiMetJet  & \checkmark & \checkmark & \checkmark \\
\minDphiMettau & \checkmark & \checkmark & \checkmark \\
\ntauTracks & \checkmark & \checkmark & \checkmark \\
\Qtauh & \checkmark & \checkmark & \checkmark \\
\sumQ & & & \checkmark \\
\bottomrule
\end{tabular}
\label{tab:input_vars}
\end{center}
\end{table}

Three event categories are used to train separate NNs, as listed in Table~\ref{tab:input_vars}. The NNs are trained using input variables based on jet and \tauh multiplicities, \bjets and \tauh identification qualities, the trigger categories, and kinematics and angular quantities. The input variables are defined below:
\begin{itemize}
\item \ntrigbuck is a trigger category index referring to which trigger has selected the event inside the category, i.e. 1 for MET, 2 for STT, and 3 for DTT.  This variable is used as an input to an NN only when the multiple event categories are merged.
\item \tauptlead is the \pt of the leading \tauh in the event.
\item \tauptsublead is the \pt of the sub-leading \tauh in the event.
\item \tauptthirdlead is the \pt of the third-leading \tauh in the event.
\item \njets is the number of reconstructed jets.
\item \ntaus is the number of \tauh candidates.
\item \HTjets is the scalar sum of \pt of all jets in the event.
\item \met is the missing transverse energy.
\item \mtautau is the invariant mass of the two leading \tauh candidates.
\item \mtaub is the invariant mass of the leading \tauh and the leading \bjet.
\item \mbbtau is the invariant mass of the system of the leading \bjets pair and the leading \tauh.
\item \mbbmet is the invariant mass of the system of the two leading \bjets and the \met.
\item \sumPCB is the sum of the pseudo-continuous $b$-tagging efficiency score for all jets in the event. Each jet score ranges from 1 to 5, corresponding to a certain range of $b$-tagging efficiencies defined by the working points: $100\%$--$85\%$, $85\%$--$77\%$, $77\%$--$70\%$, $70\%$--$60\%$, and $<60\%$.
\item \sumPCT is an integer variable (0–8) that encodes the \tauh-identification working points of up to two leading \tauh candidates in an event. The exclusive identification working points considered are Loose-not-Medium, Medium-not-Tight, and Tight. The event score is determined by the combination of working points assigned to the leading and sub-leading \tauh. %

\item \minDphiMetJet is the minimum angle in the transverse plane between the \met and any jets.
\item \minDphiMettau is the minimum angle in the transverse plane between the \met and any \tauh.
\item \ntauTracks is the number of reconstructed tracks associated with the leading \tauh.
\item \Qtauh is the charge of the leading \tauh.
\item \sumQ denotes the sum of charges of up to four \tauh candidates ordered in \pt. This variable is considered in events with two or more \tauh candidates. %
\end{itemize}

\begin{figure}
\centering
\subfloat[]{\includegraphics[width=.4\textwidth]{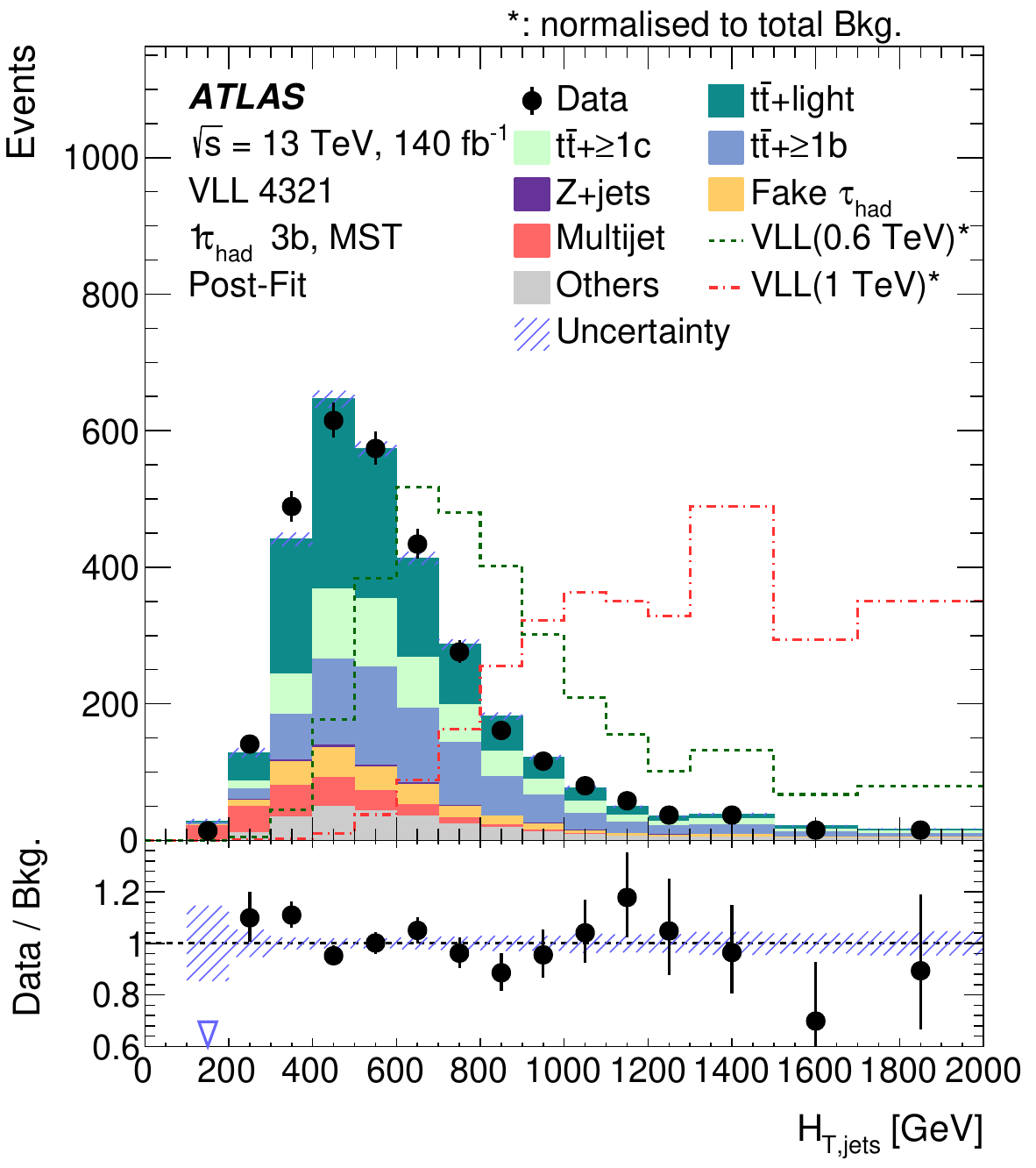}}
\subfloat[]{\includegraphics[width=.4\textwidth]{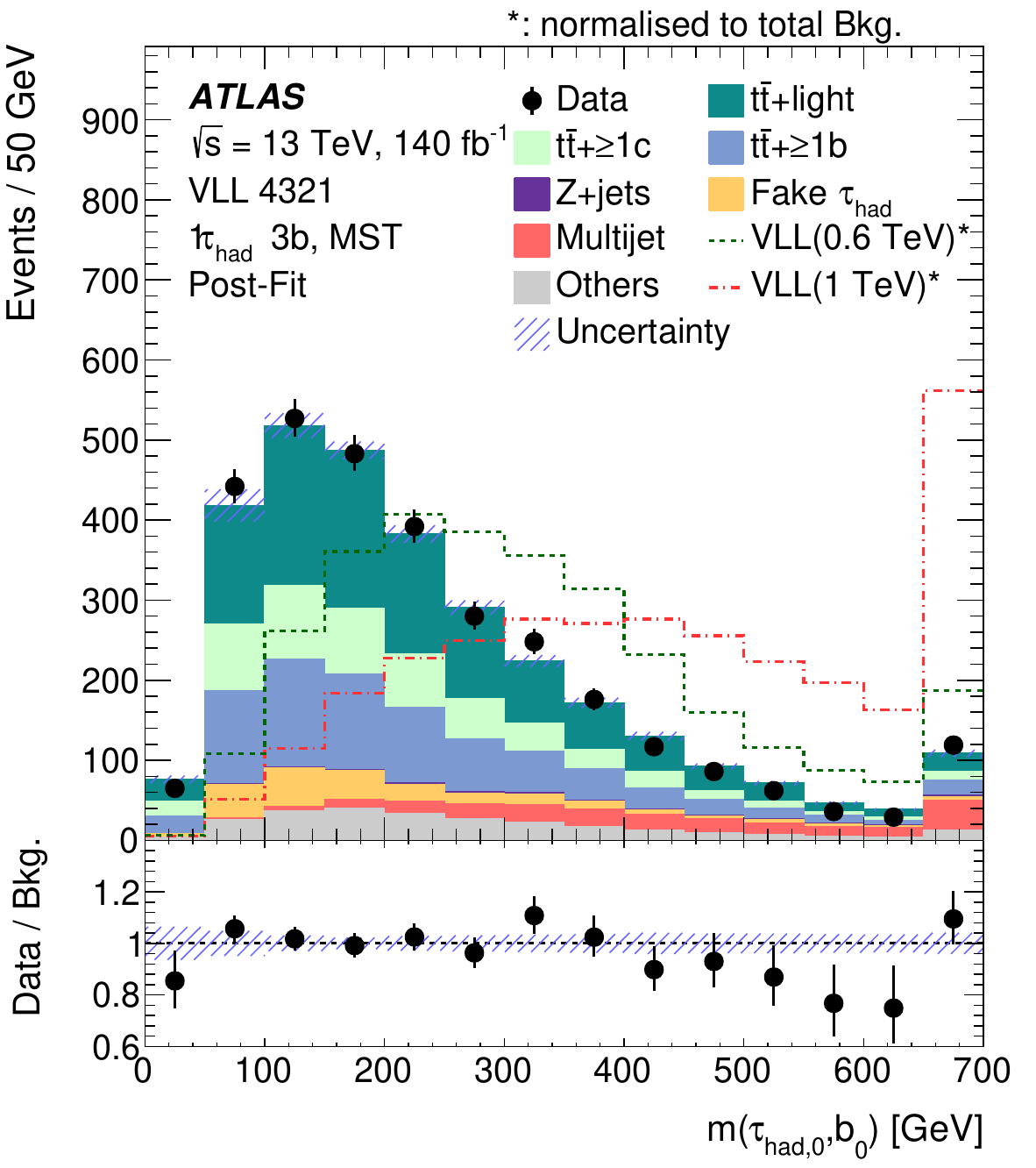}} \\
\subfloat[]{\includegraphics[width=.4\textwidth]{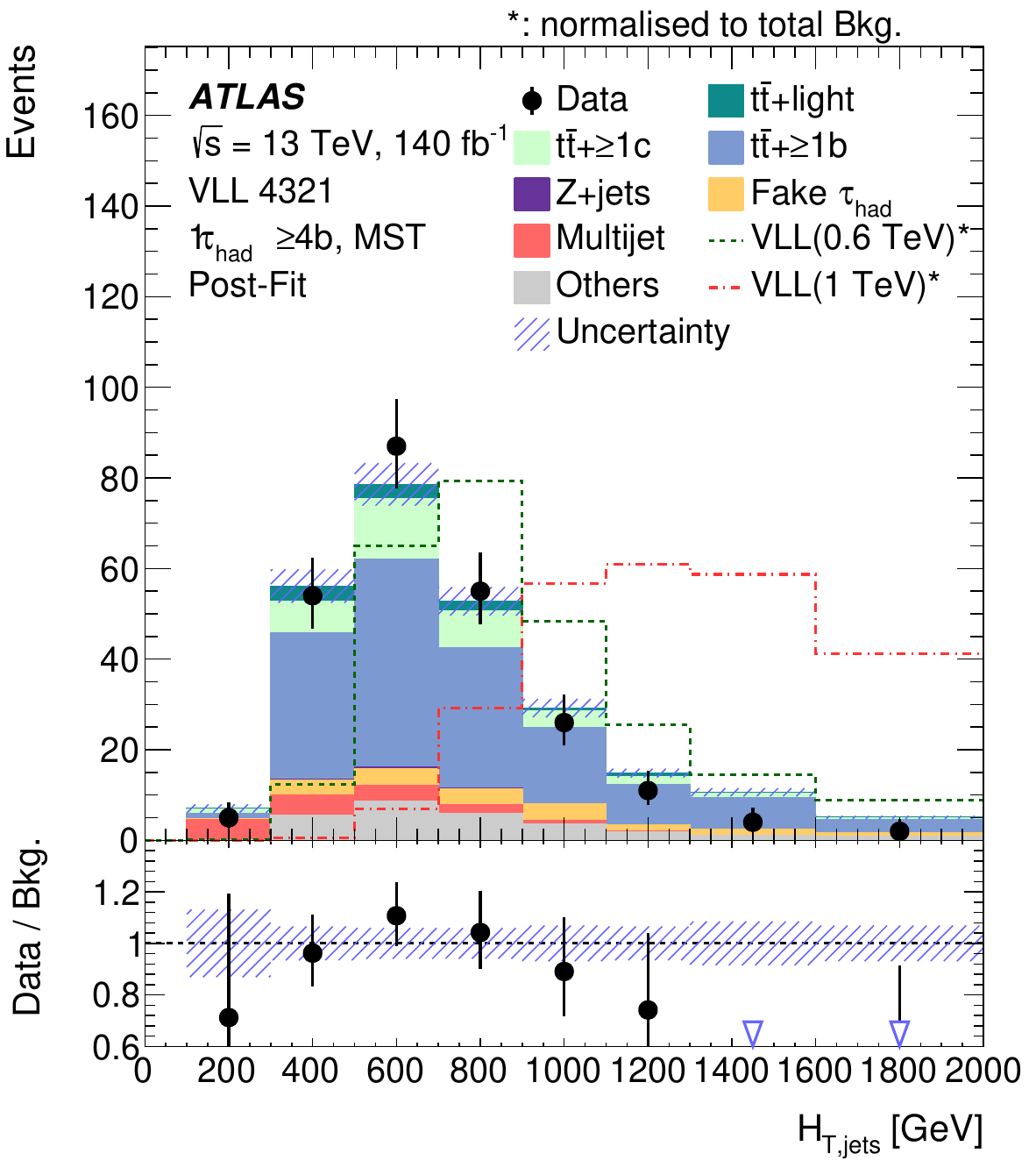}}
\subfloat[]{\includegraphics[width=.4\textwidth]{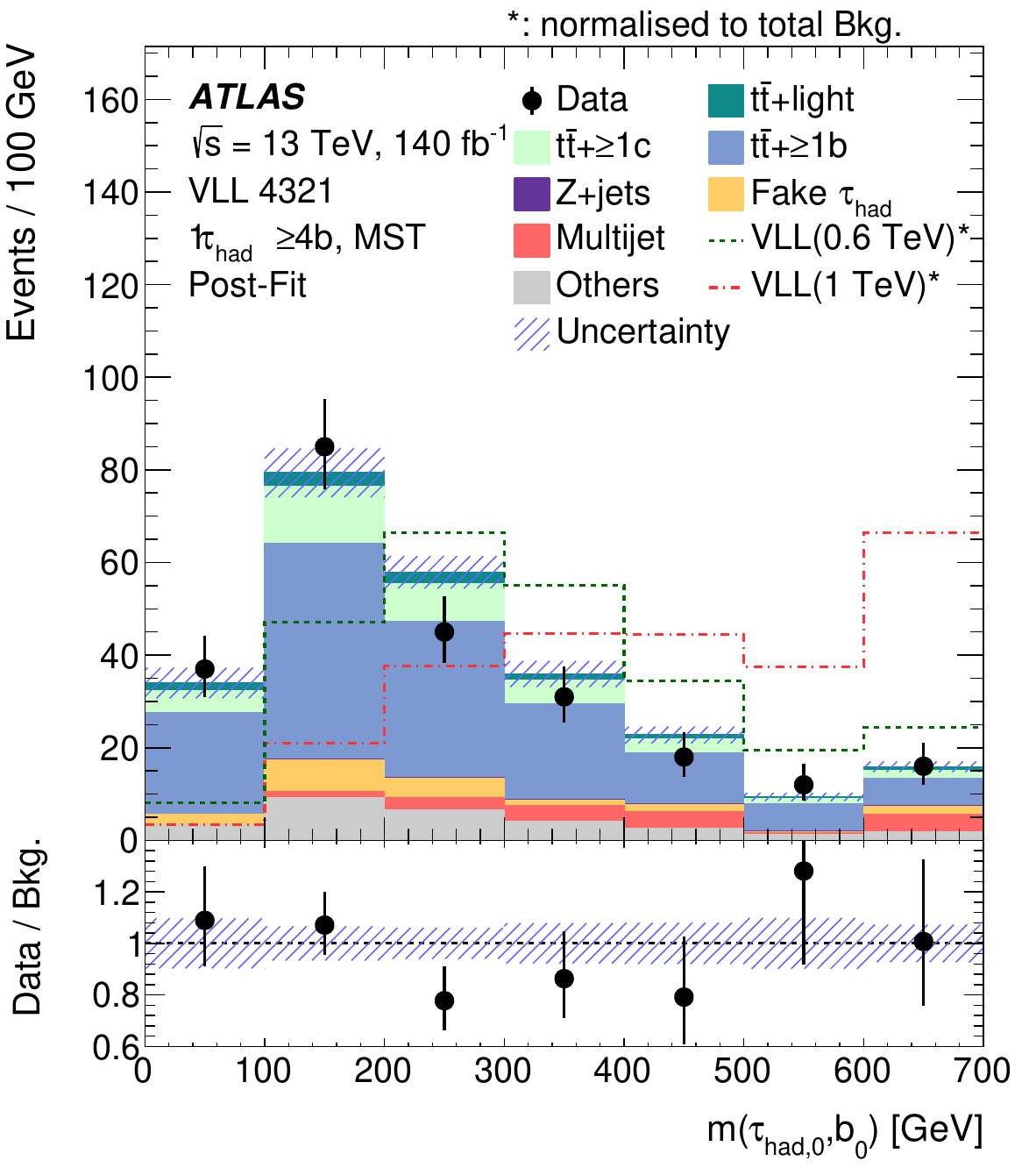}}
\caption{Post-fit background (filled histograms) and data (dots with statistical error bars) distributions of representative NN input variables in the 1$\tauh \geq\!3b$ MST signal regions; (a) \HTjets and (b) \mtaub in the 1$\tauh 3b$ MST region, (c) \HTjets and (d) \mtaub in the 1$\tauh \geq\!4b$ MST region. The normalisation and shape of the backgrounds are determined from the background-only likelihood fit to data and the ratios of the data to the sum of the predicted backgrounds are shown in the lower panels.  The hatched band indicates the combined statistical and systematic uncertainty in the total background prediction.  The expected distributions for 0.6\,\TeV\ and 1.0\,\TeV\ VLL signals, normalised to the background yields, are overlaid (dashed lines). The last bin includes the overflow.}
\label{fig:NNinputs_1tauMST_3b4b}
\end{figure}

\begin{figure}
\centering
\subfloat[]{\includegraphics[width=.4\textwidth]{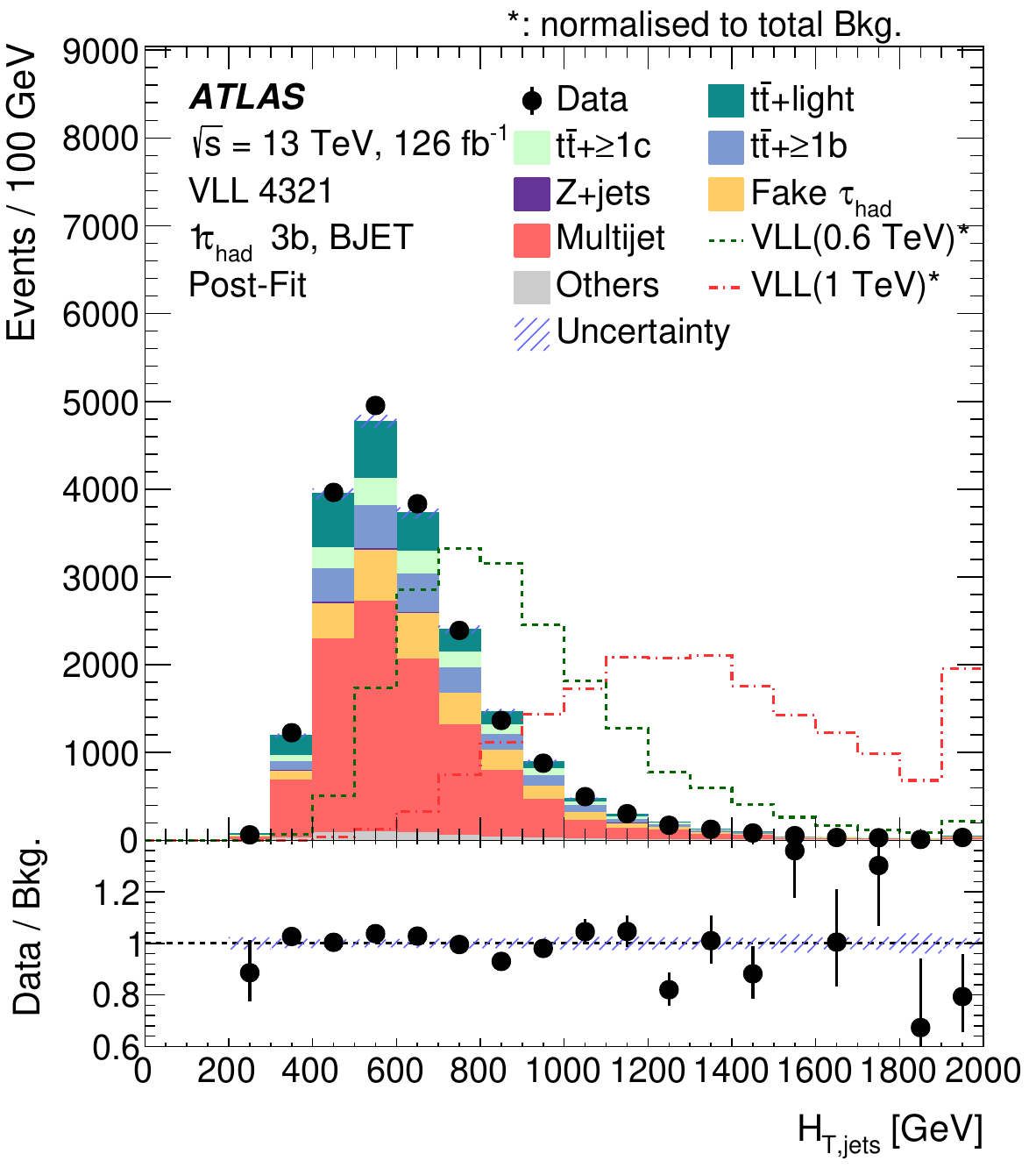}}
\subfloat[]{\includegraphics[width=.4\textwidth]{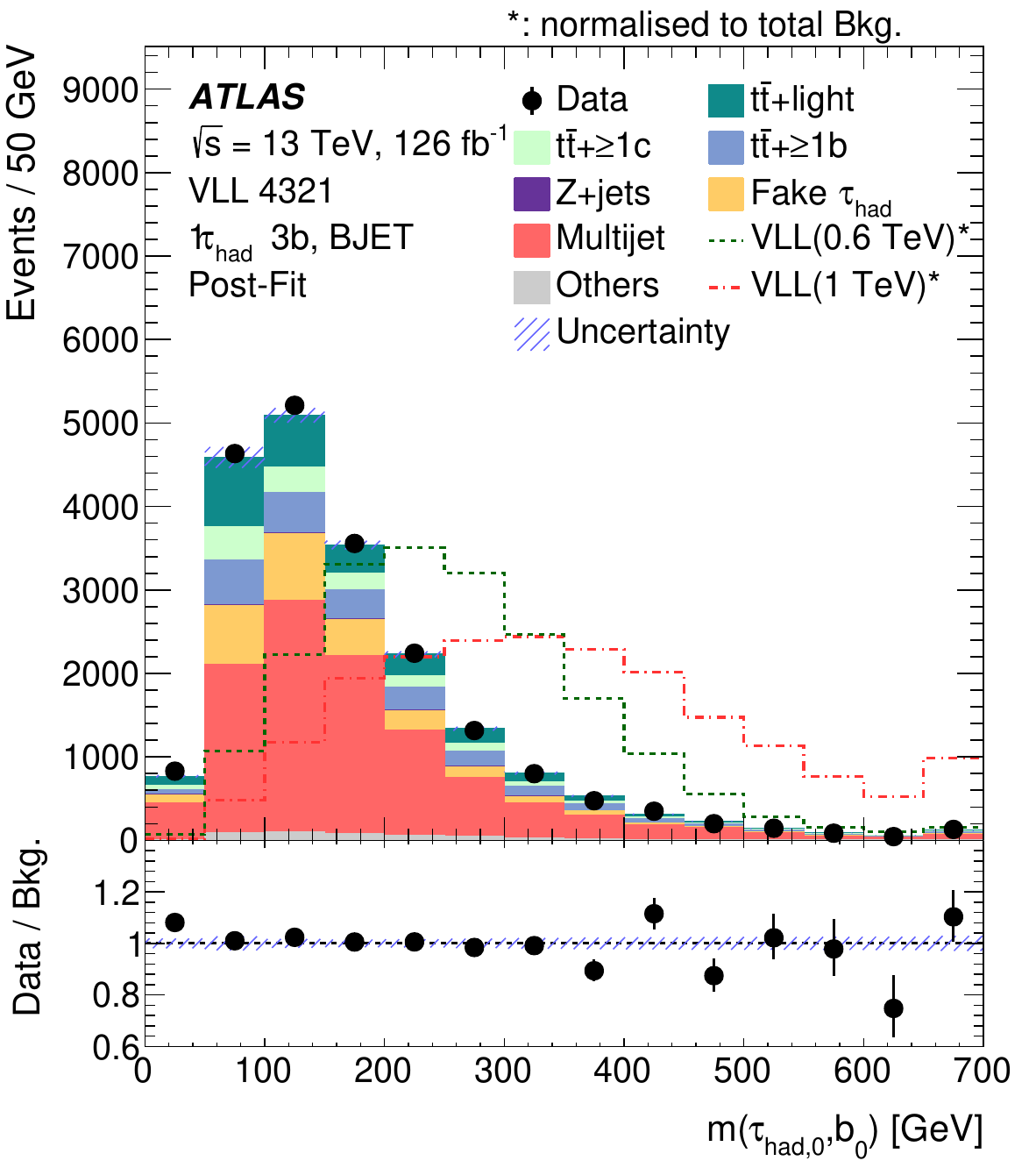}} \\
\subfloat[]{\includegraphics[width=.4\textwidth]{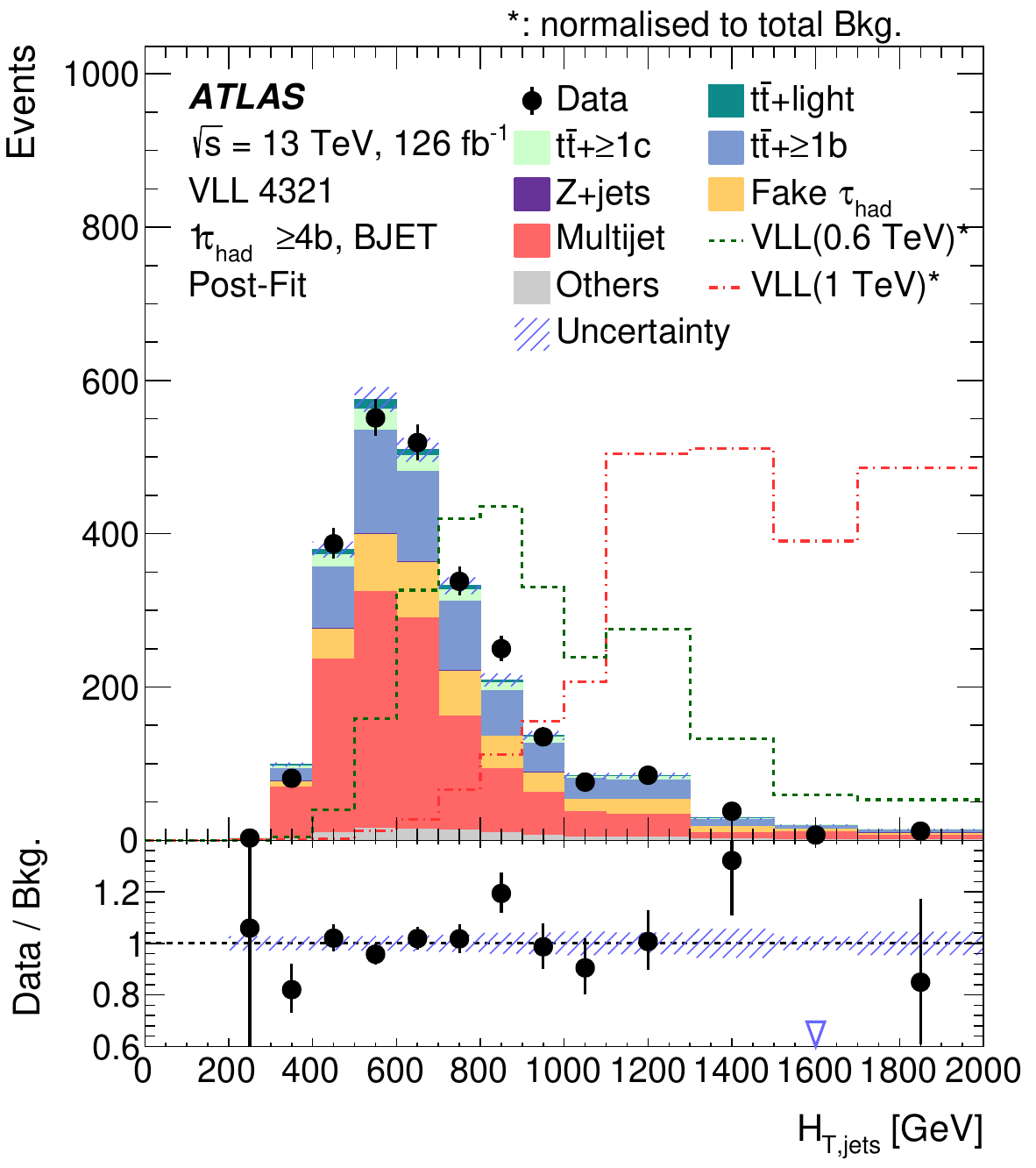}}
\subfloat[]{\includegraphics[width=.4\textwidth]{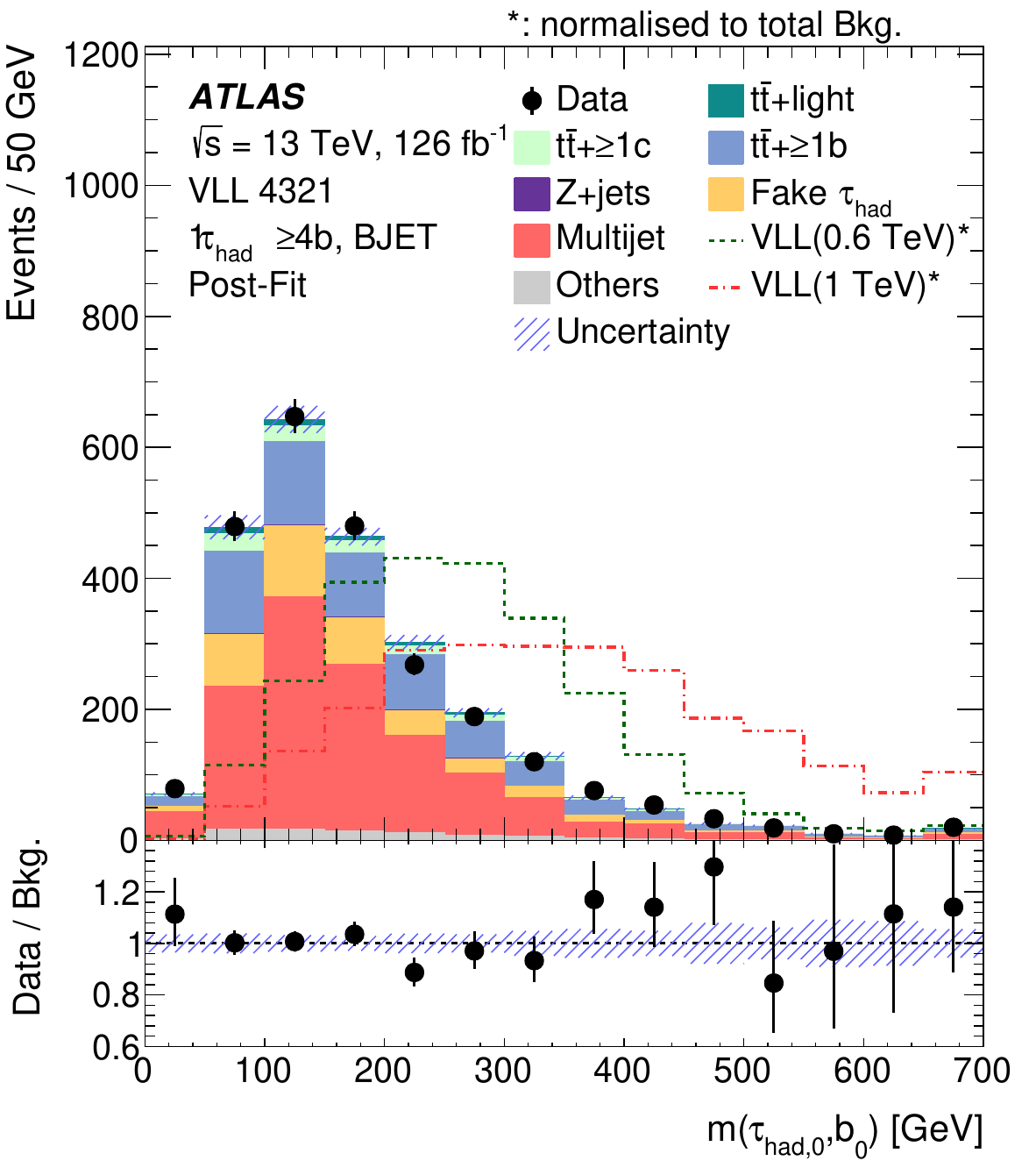}}
\caption{Post-fit background (filled histograms) and data (dots with statistical error bars) distributions of representative NN input variables in 1$\tauh \geq\!3b$ BJET signal regions; (a) \HTjets and (b) \mtaub in the 1$\tauh 3b$ BJET region, (c) \HTjets and (d) \mtaub in the 1$\tauh \geq\!4b$ BJET region. The normalisation and shape of the backgrounds are determined from the background-only likelihood fit to data and the ratios of the data to the sum of the predicted backgrounds are shown in the lower panels.  The hatched band indicates the combined statistical and systematic uncertainty in the total background prediction.  The expected distributions for 0.6\,\TeV\ and 1.0\,\TeV\ VLL signals, normalised to the background yields, are overlaid (dashed lines).  The last bin includes the overflow.}
\label{fig:NNinputs_1tauBT_3b4b}
\end{figure}

\begin{figure}
\centering
\subfloat[]{\includegraphics[width=.4\textwidth]{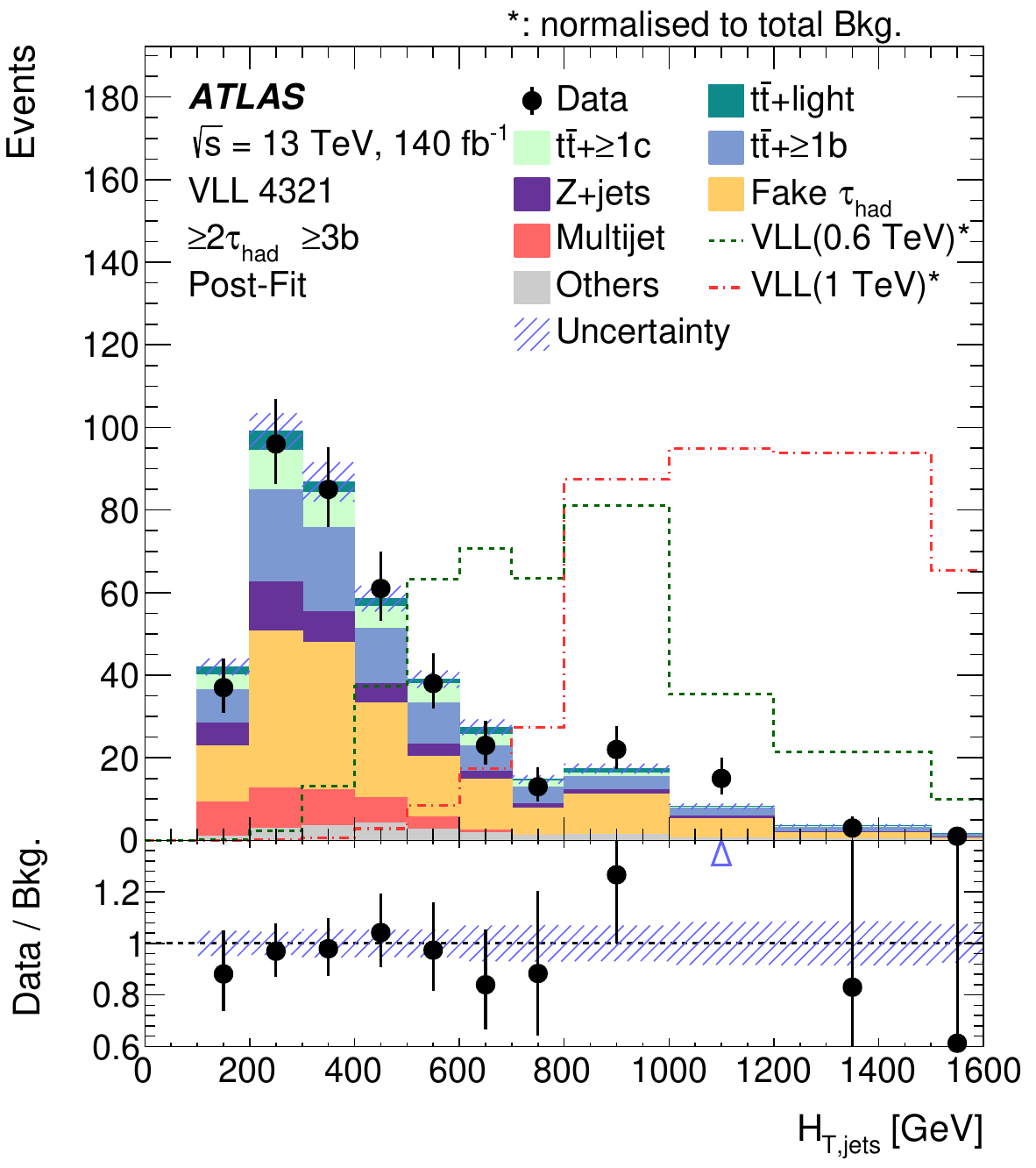}}
\subfloat[]{\includegraphics[width=.4\textwidth]{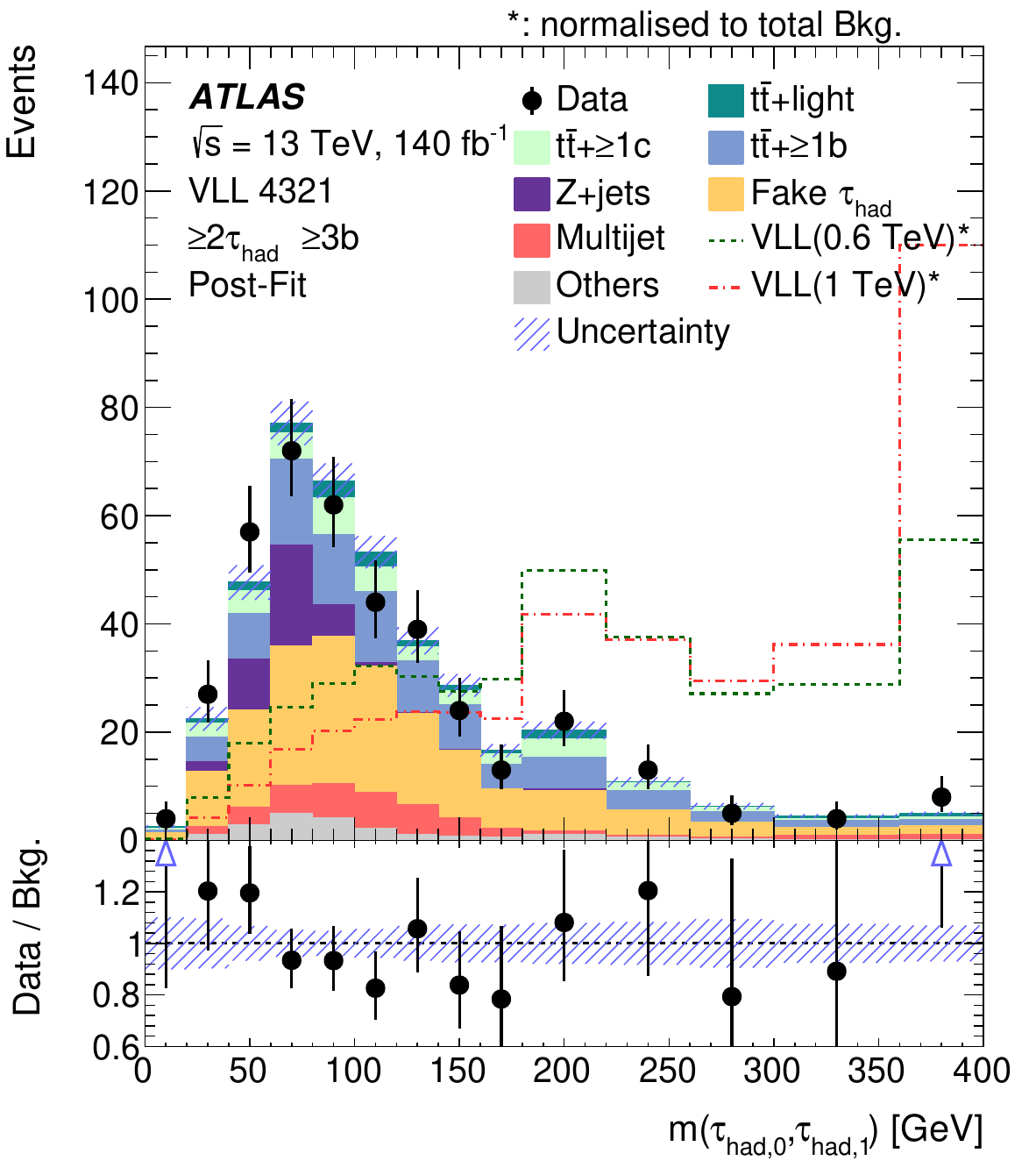}}
\caption{Post-fit background (filled histograms) and data (dots with statistical error bars) distributions of representative NN input variables in the 2$\tauh \geq\!3b$ MSDT signal region: (a) \HTjets and (b) \mtautau. The normalisation and shape of the backgrounds are determined from the background-only likelihood fit to data and the ratios of the data to the sum of the predicted backgrounds are shown in the lower panels.  The hatched band indicates the combined statistical and systematic uncertainty in the total background prediction.  The expected distributions for 0.6\,\TeV\ and 1.0\,\TeV\ VLL signals, normalised to the background yields, are overlaid (dashed lines).  The last bin includes the overflow.}
\label{fig:NNinputs_2tau}
\end{figure}

A selection of representative input distributions, after the background corrections described in Section~\ref{sec:backgrounds} and the final fit outlined in Section~\ref{sec:result}, are presented in Figures~\ref{fig:NNinputs_1tauMST_3b4b},~\ref{fig:NNinputs_1tauBT_3b4b}, and~\ref{fig:NNinputs_2tau} for the 1$\tauh \geq\!3b$ MST, the 1$\tauh \geq\!3b$ BJET, and the $\geq\!2\tauh \geq\!3b$ MSDT signal regions, respectively.  The 1$\tauh \geq\!3b$ MST region is dominated by contributions from \ttbarj processes, the 1$\tauh \geq\!3b$ BJET region has substantial contributions from the multijet production, while the $\geq\!2\tauh \geq\!3b$ MSDT region suffers from large contributions from the \ttbarj processes with real or misidentified \tauh. The input variables in these regions are consistent between the data and the SM background predictions within two standard deviations.

The NNs are trained against the main $\ttbarj$, $\Zjets$, and multijet backgrounds taking into account both the true and misidentified \tauh components with the latter corrected as described in Section~\ref{sec:backgrounds}. The jet multiplicity and additional jets flavour components are corrected in the \ttbarj sample, as described in Section~\ref{sec:backgrounds}, before employing them for training.  For the signals, the generated mass \mVLL is used as a parameterisation input in addition to the input variables described above, while in the case of the backgrounds a pseudo-mass is assigned based on the generated \mVLL value such that the training is dependent on the \mVLL mass.  Sample weights are calculated such that the signal-to-background ratio is balanced across all mass points.
Comparing the output between the training dataset and an independent testing dataset shows no sign of overtraining. The resulting NN score distributions, which peak at higher values for VLL signals than for the background processes, are used as the analysis discriminants.  The NN score distributions are binned to maximise the signal significance against the background in each SR while ensuring sufficient statistics in each bin.


%

%
%

%
\section{Background estimation}
\label{sec:backgrounds}

%

%
%
%
%
%
%
%

The composition of the SM processes varies in different signal regions, with major contributions arising from \ttbarj events with real or fake \tauh components and from multijet production. The \Zjets process enhanced in heavy-flavour component, labelled \ZHF hereafter, contributes as a small background.  As the processes with a large heavy-flavour content contributing to the analysis regions are poorly modelled, the contributions from these aforementioned processes are estimated using data-driven techniques in dedicated control regions, where the signal contamination is negligible.  Other rare processes contributing as irreducible background are estimated using MC simulations.
The modelling of background processes is validated in the five SRs with $1\tauh$ and $\geq\!2\tauh$ event selections using the lower bins of NN score variable distributions.
The data-driven estimation of the \ttbarj, \Zjets and multijet background processes are described in the following sections.

\subsection{\ttbarj background}
\label{sec:ttbar}
The \ttbarj processes are an important background in this search, in particular the processes with at least one additional heavy-flavour jet. The jets mis-tagged as \bjets further enhance the background rate.  The state-of-the-art predictions of \ttbarj processes at NLO in QCD suffer from large uncertainties in the phase space with multiple jets. The kinematics of the top quark as predicted in the NNLO+NNLL QCD and NLO EW calculation~\cite{Czakon_2017} better describe the measurements as compared with the predictions at NLO. However, these calculations are inclusive in additional jet flavour, and may not describe the data well at high jet multiplicities.  It is therefore crucial and common to constrain the modelling of \ttbarj process using data. The \ttbar process with both top quarks decaying semileptonically (into $e$ or $\mu$) provides a much cleaner final state compared to that with hadronic decays (into hadrons or \tauh). A scale factor region selected with an electron, a muon and two or more jets is used to first derive \ttbar modelling corrections for the reconstructed jet multiplicity spectrum and for the composition of additional flavoured jets in the \ttbarj sample. The final normalisation of \ttbarj with heavy-flavour components in the signal regions is determined from the simultaneous fit to data in the signal and control regions selected with at least one \tauh and no light leptons. The multiple steps involved in deriving the corrections are described in the following.

\begin{table}[b]
\caption{Summary of the selections which define the regions used to determine the reducible background shape correction to the \ttbarj, \Zjets and fake \tauh processes, and to validate the fake \tauh modelling.  The `SF' (`VR') stands for the scale factor determination (validation) region. The $\mathrm{`OS'}$ refers to an opposite-charge light-lepton pair, $m_{\ell\ell}$ denotes the invariant mass of two light leptons, and $m_Z$ is the $Z$ boson mass. The two columns under the fake \tauh SF category refer to the regions enriched in \ttbarj and \Zjets, in respective order. The $\mathrm{`SS'}$ refers to the same-charge \tauh pair.}
\begin{center}
\resizebox{1\linewidth}{!}{%
\begin{tabular}{l|c|c|c|c|c|c|c}
\toprule
Selections & \multicolumn{1}{c|}{\ttbarj} & \multicolumn{1}{c|}{\Zjets}  & \multicolumn{3}{c|}{Fake \tauh} & \multicolumn{2}{c}{Multijet} \\
& \multicolumn{1}{c|}{SF} & \multicolumn{1}{c|}{SF}  & \multicolumn{2}{c|}{SF} & \multicolumn{1}{c|}{VR} &  \multicolumn{2}{c}{SF} \\
\midrule
Triggers  & \multicolumn{4}{c|}{SLT $\cup$ DLT $\cup$ MET} & \multicolumn{1}{c|}{MET $||$ BJET } & STT $||$ BJET & STT $||$ DTT\\
Number of $b$-tags  & \multicolumn{2}{c|}{---} & \multicolumn{2}{c|}{1 or 2 $@85\%$} & \multicolumn{1}{c|}{1 or 2 $@77\%$} & 1$@77\%$ & 0,1,2, $\ge$3$@77\%$ \\
Number of jets  & \multicolumn{5}{c|}{$\geq$2} & $\ge$1 & $\ge$3 \\
Number of \tauh & \multicolumn{2}{c|}{0} & \multicolumn{3}{c|}{1} & 1 & 2(SS) \\
Number of $e/\mu$  & \multicolumn{1}{l|}{$e\mu\mathrm{OS}$}  & \multicolumn{1}{l|}{$ee\mathrm{OS}$ or $\mu\mu\mathrm{OS}$}  & \multicolumn{1}{l|}{$e\mu\mathrm{OS}$} & \multicolumn{1}{l|}{$ee\mathrm{OS}$ or $\mu\mu\mathrm{OS}$} & \multicolumn{1}{l|}{1 (same charge as \tauh)} & \multicolumn{2}{c}{0}\\
Lepton \pt\ {[}\GeV{]} & \multicolumn{5}{c|}{(20, 20)} & \multicolumn{2}{c}{---} \\
Lepton identification   & \multicolumn{5}{c|}{Electron: TightLH} & \multicolumn{2}{c}{---}  \\
& \multicolumn{5}{c|}{Muon: Medium} & \multicolumn{2}{c}{---} \\
$m_{\ell\ell}$ {[}\GeV{]} & \multicolumn{1}{c|}{---}  & \multicolumn{3}{c|}{$>15$} & \multicolumn{1}{c|}{---} & \multicolumn{2}{c}{---}\\
$|m_{\ell\ell} - m_{Z}| < 10$ {[}\GeV{]}  & \multicolumn{1}{c|}{---}  & \multicolumn{1}{c|}{yes} & \multicolumn{1}{c|}{---} & \multicolumn{1}{c|}{yes} & \multicolumn{1}{c|}{---} & \multicolumn{2}{c}{---} \\
\bottomrule
\end{tabular}
}
\label{tab:ttbarZjetsFakeTauCRs}
\end{center}
\end{table}

First, the MC simulated \ttbar events are reweighted using an iterative recursive procedure to correct the parton level distributions of the top quark \pt, invariant mass of the \ttbar system, and the \pt of the \ttbar system to those predicted by the NNLO+NNLL QCD and NLO EW corrections~\cite{Czakon_2017}.  Second, the corrections for the jet multiplicity are determined for the reweighted \ttbar sample using a scale factor region as described in the second column of Table~\ref{tab:ttbarZjetsFakeTauCRs}. %
The background contamination from other processes is less than $11\%$ in this region. The corrections are defined as the ratio of data events after subtracting all non-\ttbar background to the predicted \ttbar events in each jet multiplicity bin.  They range from $5\%$ for events with two jets to about $30\%$ for 10 or more jets.

In the next step, the MC events in the above mentioned scale factor region are further categorised (following the definitions described in Ref.~\cite{EXOT-2017-11,HIGG-2020-23}) depending on the truth flavour of additional QCD jets produced with \ttbar. If there is at least one additional \bjet, the event is classified as \ttb; if there is no additional \bjet, but an additional \cjet is present, it is classified as \ttc, or else, as \ttl. Templates are created for the three components of \ttbar in terms of a discriminating variable defined as the sum of $b$-tagging score (\sumPCB), where each jet score refers to a range of average $b$-tagging efficiencies. A likelihood fit is performed to data in the scale factor determination region for the estimation of the normalisation factors of the three \ttbarj categories.  These estimated normalisation factors are used as initial corrections to the \ttbar MC events in all analysis regions. %

\begin{table}[t]
\caption{Summary of the selection requirements for each of the seven control regions. Each control region is labelled depending on the combination of triggers, the multiplicity of \tauh candidates, number of \bjets, and other kinematic requirements. The $E_{\mathrm{T, lo}}^{\mathrm{miss}}$ ($E_{\mathrm{T, hi}}^{\mathrm{miss}}$) refers to the low (high) \met regions. %
The $m_{\tau\tau, \mathrm{lo}}^{\mathrm{coll}}$ ($m_{\tau\tau, \mathrm{hi}}^{\mathrm{coll}}$) refers to regions with low (high) range of the invariant mass variable $m_{\tau\tau}^{\mathrm{coll}}$ as described in Section~\ref{sec:Zjets}. The $\mathrm{`OS'}$ refers to opposite charge \tauh pair. The \ZLF (\ZHF) refers to the region enriched in $Z$ boson events without (with) additional $b$-tagged jets.}
\begin{center}
\resizebox{1\linewidth}{!}{%
\begin{tabular}{c|c|c|c|c|c|c|c}\toprule
Control region & \ttl & \multicolumn{3}{c|}{Multijet}  & \ZLF & \ZHF & Multijet \\
& 1$\tauh 2b$ $E_{\mathrm{T, hi}}^{\mathrm{miss}}$   & 1$\tauh 2b$ $E_{\mathrm{T, low}}^{\mathrm{miss}}$  & 1$\tauh 2b$ $E_{\mathrm{T, low}}^{\mathrm{miss}}$  & 2$\tauh 0b$ $m_{\tau\tau, \mathrm{hi}}^{\mathrm{coll}}$  & 2$\tauh 0b$ $m_{\tau\tau, \mathrm{low}}^{\mathrm{coll}}$ & 2$\tauh 1b$ $m_{\tau\tau, \mathrm{low}}^{\mathrm{coll}}$ & 2$\tauh 2b$  \\
& MST &  MST & BJET &   &  &  &   \\
\midrule
Triggers & \multicolumn{2}{c|}{MET $\cup$ STT} & BJET & \multicolumn{4}{c}{MET $\cup$ STT $\cup$ DTT} \\
Number of \tauh & \multicolumn{3}{c|}{$1$} & \multicolumn{4}{c}{$2$ (OS) or $\geq\!3$ (STT, DTT), or $\geq\!2$ (MET)} \\
Number of $e/\mu$ & \multicolumn{7}{c}{0}\\
Number of jets & \multicolumn{3}{c|}{$\geq\!4$}  & \multicolumn{4}{c}{$\geq \!3$} \\
Number of $b$-tags & \multicolumn{3}{c|}{$2@77\%$} & \multicolumn{2}{c|}{$0@77\%$} & $1@77\%$ & $2@77\%$  \\

$m_{\tau\tau}^{\text{coll}}$ & \multicolumn{3}{c|}{-} & > 120 GeV & \multicolumn{2}{c|}{< 120 GeV} & - \\
$E_{T}^{\text{miss}}$  & $>$ 100 GeV & \multicolumn{2}{c|}{$<$ 100 GeV} & \multicolumn{4}{c}{-} \\

Fit variable & \multicolumn{6}{c|}{Event yield} & \sumPCT \\  \bottomrule
\end{tabular}
}
\label{tab:taucrs}
\end{center}
\end{table}

Finally, the normalisation of \ttb, \ttc and \ttl processes is determined in the simultaneous likelihood fit to data in the signal and control regions consisting of one or more \tauh but depleted in light leptons. The fit setup is outlined in Section~\ref{sec:result}. The control region, as summarised in Table~\ref{tab:taucrs}, is characterised by events with one \tauh, at least four jets, of which two are $b$-tagged at the $77\%$ efficiency working point~\cite{FTAG-2019-07}, and \met above $\SI{100}{\GeV}$.  Events are required to satisfy the MET or STT triggers. The normalisation of \ttb, \ttc and \ttl categories in the alternative \ttbarj samples used to estimate systematic uncertainties are scaled to the estimation of \ttb, \ttc and \ttl yields in the nominal \ttbar sample, such that only the shape uncertainties are taken into account in the final fit.  %
The measured normalisation factors (including the initial normalisation corrections estimated from scale factor regions selected with light leptons) are $0.93\pm 0.10$, $1.27\pm 0.53$, and $1.41\pm 0.32$ for \ttl, \ttc and \ttb components, respectively, where the quoted error includes all uncertainties.  The correlation factor between the \ttl (\ttc) and \ttb normalisation estimators is about $50\%$ ($60\%$), whereas the normalisation factors of \ttl and \ttc have an anti-correlation factor of $60\%$. The measured values refer to the case where the NN score is evaluated for data and the SM background corresponding to a VLL mass of 1~TeV, they do not change significantly for the other mass values.

\subsection{\Zjets background}
\label{sec:Zjets}
The corrections to the MC simulated \Zjets events are derived in terms of the reconstructed jet multiplicity using the region consisting of two opposite-charge same-flavour light leptons, each with \pt above $\SI{20}{\GeV}$ and with an invariant mass within $\SI{10}{\GeV}$ of the $Z$ boson mass. The number of jets exhibits good consistency between the data and MC at low multiplicity of up to five jets. However, it requires a correction at the level of 25\% when it increases to 10 jets.

After the scale factors are applied to the \Zjets MC events correcting for the reconstructed jet multiplicity, the normalisation of \Zjets is determined using events containing at least two \tauh and no light leptons.  The \Zjets MC events are split in two orthogonal components, denoted as as \ZLF and \ZHF, depending on the truth flavour of the jets produced in association with the $Z$ boson. The \ZHF component consists of events with at least one reconstructed jet matched to a truth \bjet or \cjet, and all remaining events are classified as \ZLF. %
Two orthogonal control regions selected with events containing two \tauh candidates and at least three jets are summarised in Table~\ref{tab:taucrs}. One region is dominated with \ZLF events with the presence of no $b$-tagged jet and the other, having one $b$-tagged jet, is enriched in \ZHF events.
A mass variable  $m_{\tau\tau}^{\mathrm{coll}}$ is constructed under the approximation~\cite{CollinearApprox} that the neutrino is collinear with the visible decay products of \tauh and is the only source of \met in the event. It is defined as $m_{\tauh\tauh}/\sqrt{x_{\tau_{\mathrm{had},0}}x_{\tau_{\mathrm{had},1}}}$, where $m_{\tauh\tauh}$ is the invariant mass of the \tauh pair and $x_{\tau_{\mathrm{had},0}}$ ($x_{\tau_{\mathrm{had},1}}$) is the fraction of the original \pt of the leading (sub-leading) reconstructed \tauh that is carried by its visible decay products. The original \pt refers to the total estimated \pt of \tauh, obtained by combining the visible decay products with a component of the \met associated with neutrinos from tau decay.
The variable $x_{\tau_{\mathrm{had},0/1}}$ expressed in terms of the kinematic variables of \tauh and \met is given by

\begin{equation*}
x_{\tau_{\mathrm{had},0/1}} = \frac{p_{\mathrm{T},\tau_{\mathrm{had},0/1}}}{p_{\mathrm{T},\tau_{\mathrm{had},0/1}} + E_{T}^{\mathrm{miss}} (\sin(\phi_{E_{T}^{\mathrm{miss}}}) - \sin(\phi_{\tau_{\mathrm{had},1/0}})) / (\sin(\phi_{\tau_{\mathrm{had},0/1}}) - \sin(\phi_{\tau_{\mathrm{had},1/0}}))},
\end{equation*}

where $\phi_{E_{T}^{\mathrm{miss}}}$ is the azimuth coordinate of the missing transverse momentum, and $\phi_{\tau_{\mathrm{had},0}}$ ($\phi_{\tau_{\mathrm{had},1}}$) is the azimuth coordinate of the leading (sub-leading) \tauh. %
The \Zjets enriched region is selected with the requirement of $m_{\tau\tau}^{\mathrm{coll}}$ below $\SI{120}{\GeV}$.  The measured values of normalisation factors for \ZLF and \ZHF determined from the fit to data in the signal and control regions considering the background-only hypothesis and an \mVLL of 1~TeV for the NN score distribution are $1.23\pm 0.11$ and $1.13 \pm 0.10$, respectively, where the quoted error includes all uncertainties. The fitted values of the \ZLF and \ZHF normalisation factors have a correlation factor of about $50\%$.

\subsection{Fake \tauh background in non-multijet production}
\label{sec:faketau}
The \ttbarj process also contributes as a non-negligible background when at least one of the jets is misidentified as a \tauh. This is one of the important contributions in the $\geq\!2\tauh\geq\!3b$ MSDT signal region. The background is estimated using a semi-data-driven approach, where the predictions from the MC simulations are taken as baseline, and corrections are applied to the MC simulated events containing one or more fake \tauh depending on each fake \tauh \pt and number of associated tracks.  The corrections are derived using a dedicated region enriched in \ttbarj events consisting of one \tauh with a minimum \pt of $\SI{20}{\GeV}$, as summarised in Table~\ref{tab:ttbarZjetsFakeTauCRs}.  As discussed in Section~\ref{sec:ttbar}, the \ttbarj MC events are first corrected for the mis-modelling in the top quark kinematics at the truth level, the reconstructed jet multiplicity mis-modelling, and for the relative contributions of \ttb, \ttc and \ttl processes.  %
The scale factors are defined as the ratio of observed yields in the control region after subtracting the non-fake-\tauh background to the simulated fake \tauh predictions in bins of the \pt and associated track multiplicity (1-prong or 3-prong) of the fake \tauh.  They are also separated in terms of exclusive efficiencies of \tauh identification.  The fake \tauh with \pt $\leq 100$~GeV have scale factors ranging from about 0.7 to 1.2 (0.8 to 1.2) for 1-prong (3-prong) fake \tauh, depending on the \pt and the identification working point.  The relative statistical uncertainties in these corrections range from $10\%$ to $30\%$. The scale factors for fake \tauh with \pt $> 100$~GeV range between 0.5 and 0.9 (0.9 and 1.2) for 1-prong (3-prong) fake \tauh, with statistical uncertainties of up to $30\%$ ($20\%$-$60\%$). The fake \tauh rate depends on whether \tauh originates from a quark jet or a gluon jet. Any mis-modelling depending on the fake \tauh origin from quark- or gluon-jets is accounted for as systematic uncertainties in the scale factors.  As the composition of quark and gluon jets is different between the \ttbarj and \Zjets processes, the scale factors are similarly derived using an alternative determination region dominated by \Zjets events, as listed in the right sub-column of the fake \tauh SF column in Table~\ref{tab:ttbarZjetsFakeTauCRs}. These alternative SFs are consistent with the nominal within $10\%$. They are used as systematic variations to cover for potential differences in the fake \tauh origin across the control regions and the signal regions.

The estimated scale factors are validated in an independent validation region selected with one light lepton and one \tauh with the same charge, as summarised in Table~\ref{tab:ttbarZjetsFakeTauCRs}. Events selected with the MET or BJET trigger category, at least two jets (of which one or two are $b$-tagged) and exactly one light lepton include fake \tauh contributions arising from various processes, such as \ttbarj, \Wjets and \Zjets.  These processes lead to different jet flavor composition faking \tauh. The data show consistency with the modelling of the fake \tauh backgrounds. %
The validation of fake \tauh scale factors is further demonstrated in the distributions of \HTjets and \mtautau, as shown in Figure~\ref{fig:NNinputs_2tau}, in the 2$\tauh$ $\geq\!3b$ region where the fake \tauh background is the dominant contribution.

\subsection{Multijet production background}
The multijet background originating from QCD processes is estimated using two data-driven methods: for channels with a single \tauh the fake factor method~\cite{EGAM-2019-01} is used, while for two or more \tauh channels the charge symmetry in the multijet production is exploited. Details on the selection regions used for the calculation of the scale factors can be found in Table \ref{tab:ttbarZjetsFakeTauCRs}.

The multijet QCD background estimate in the single \tauh selection uses the fake factor method and relies on Loose, Loose-not-Tight and Tight \tauh identification working points, where Loose-not-Tight refers to the \tauh identification passing the Loose but failing the Tight definition.  Two independent control regions are used to estimate this background: 1) a region with Loose-not-Tight \tauh, but otherwise the same event selection as the 1\tauh SRs described in Table~\ref{tab:tauonly_srs}; 2) a region with Loose \tauh and with one $b$-tagged jet, as described in the second-to-last column of Table~\ref{tab:ttbarZjetsFakeTauCRs}.  A transfer factor parameterised in $p_{\mathrm{T}}^{\tau_{\mathrm{had}}}$, defined as $\varepsilon_{\mathrm{QCD}}(p_{\mathrm{T}}^{\tau_{\mathrm{had}}})$ = $\tilde{N}^{\mathrm{T}}_{\mathrm{QCD}}(p_{\mathrm{T}}^{\tau_{\mathrm{had}}})/\tilde{N}^{\mathrm{L}}_{\mathrm{QCD}}(p_{\mathrm{T}}^{\tau_{\mathrm{had}}})$, is derived using the region 2), where $\tilde{N}^{\mathrm{T}}_{\mathrm{QCD}}(p_{\mathrm{T}}^{\tau_{\mathrm{had}}})$ and $\tilde{N}^{\mathrm{L}}_{\mathrm{QCD}}(p_{\mathrm{T}}^{\tau_{\mathrm{had}}})$ represent the number of Tight and Loose data events, after subtracting the non-QCD background.  The $\varepsilon_{\mathrm{QCD}}(p_{\mathrm{T}}^{\tau_{\mathrm{had}}})$ is used to evaluate the number of QCD data events, $N^{\mathrm{T}}_{\mathrm{QCD}}(p_{\mathrm{T}}^{\tau_{\mathrm{had}}})$, in the $1\tauh \geq3b$ SRs with Tight selection:
\begin{equation*}
N^{\mathrm{T}}_{QCD}(p_{\mathrm{T}}^{\tau_{\mathrm{had}}}) = \frac{\varepsilon_{\mathrm{QCD}}(p_{\mathrm{T}}^{\tau_{\mathrm{had}}})}{1-\varepsilon_{\mathrm{QCD}}(p_{\mathrm{T}}^{\tau_{\mathrm{had}}})} \left[N^{\mathrm{T-L}}_{\mathrm{data}}(p_{\mathrm{T}}^{\tau_{\mathrm{had}}}) -  N^{\mathrm{T-L}}_{\mathrm{nonQCD}}(p_{\mathrm{T}}^{\tau_{\mathrm{had}}})\right]
\end{equation*}

where $N^{\mathrm{T-L}}_{\mathrm{data}}(p_{\mathrm{T}}^{\tau_{\mathrm{had}}})$ refers to the observed event count passing Loose-not-Tight \tauh in region 1) and $N^{\mathrm{T-L}}_{\mathrm{nonQCD}}(p_{\mathrm{T}}^{\tau_{\mathrm{had}}})$ is the non-QCD event yield from the simulation in the same selection.  This method uses the fact that the number of data events in the Loose region equals the sum of events in the Loose-not-Tight and Tight regions.  The parameterised $\varepsilon_{\mathrm{QCD}}$ estimate is done in binned $p_{\mathrm{T}}^{\tau_{\mathrm{had}}}$ distributions separately for 1- and 3-prong \tauh in each of the STT and BJET selections. A shifted inverse power-law (linear) fit function is used for the 1-prong (3-prong) \tauh.

The estimated multijet background, derived from the parameterised fit functions is validated using events containing single Tight-\tauh, one \bjet, and at least four jets in the MST and BJET trigger categories.  The data show good agreement with the predicted \tauh \pt and jet multiplicity distributions, both before and after the fit.  Furthermore, Figure~\ref{fig:NNinputs_1tauBT_3b4b} shows that the multijet background in the $\geq 3b$ region is well-modelled across key kinematic variables used in NN training.
Channels containing two oppositely charged \tauh candidates use the charge symmetry method to obtain the multijet background. In this method, the multijet yields in regions with two opposite-charge \tauh and two same-charge \tauh are assumed to be equal, $N_{\mathrm{QCD}}^{2\tauh \mathrm{OS}} = N_{\mathrm{QCD}}^{2\tauh \mathrm{SS}}$.
The number of events in the $2\tauh \mathrm{SS}$ selection is obtained from the data subtracted by the non-QCD MC events, for which corrections are applied as highlighted in \cref{sec:ttbar,sec:Zjets,sec:faketau}.

The final normalisations of multijet background in the signal regions are determined in the simultaneous fit to data in the signal and control regions, as summarised in Tables~\ref{tab:tauonly_srs} and ~\ref{tab:taucrs}, respectively.  The measured values of the three normalisation factors in the $1\tauh \geq\!3b$ MST, $1\tauh\geq\!3b$ BJET, and $2\tauh \geq\!3b$ MSDT regions are $0.97\pm0.20$, $0.89\pm0.04$, $1.02\pm0.04$, respectively, with correlations of less than $20\%$ among the corresponding fitted parameters. These values refer to the scenario where the NN score distributions in a signal region are evaluated considering the signal generated mass of $1$ TeV.  %
The multijet background estimate is validated using regions selected with $0b$, $1b$, $2b$ and $\ge3b$ and two opposite-charge \tauh for the STT and DTT trigger categories, with an additional requirement of at least three jets in the region, and good agreement is found in all regions.


%

%
\section{Systematic uncertainties}
\label{sec:systematics}

%
%
Various sources of systematic uncertainties are considered in the analysis. These include the uncertainties from the reconstructed objects and from the theory modelling of various physics processes that can impact the estimated signal and background rates and the shape of the fitted distributions in multiple event categories. The uncertainty on the integrated luminosity is \SI{0.83}{\%}~\cite{DAPR-2021-01}.

The uncertainties in the scale factors applied to correct for the efficiency differences between the data and MC simulations in \tauh~\cite{ATLAS-CONF-2017-061}, \bjet and \met triggers, \tauh reconstruction and identification~\cite{PERF-2013-06, ATL-PHYS-PUB-2019-033,ATLAS-CONF-2017-029}, jet reconstruction and vertex tagging~\cite{PERF-2014-03,ATLAS-CONF-2017-065}, $b$-tagged jet identification~\cite{FTAG-2018-01}, and for the pile-up modelling~\cite{ATLAS-CONF-2012-042} in the simulated events as discussed in Section~\ref{sec:objects}, are accounted for and they have a minor impact on the results.  Furthermore, uncertainties are considered for the energy scale and resolution of the jets~\cite{JETM-2018-05}, \tauh~\cite{ATL-PHYS-PUB-2019-033} and \met~\cite{JETM-2020-03}, and are found to be negligible.  All of these experimental uncertainties are treated as correlated among the signal and background processes.

In order to estimate the effect of higher-order corrections in the calculations, the impact of variations in the QCD renormalisation ($\mu_{R}$) and factorisation ($\mu_{F}$) scales, with respect to the nominal scales, as well as the uncertainties in the PDFs are evaluated on the VLL signal acceptance in the signal regions.

Modelling uncertainties associated with the \ttbarj processes are evaluated by comparing the fit distributions from the nominal MC \ttbar samples with those from the alternative MC generator predictions, as listed in Table~\ref{tab:mcconfig}. For these uncertainties, the nuisance parameters (NPs) representing the \ttc and \ttl processes, as well as those representing various \ttb processes ($\ttoneb$ with the additional \bjet containing one $b$-hadron, $\ttoneB$ with the additional \bjet containing more than one $b$-hadrons, $\tttwob$, and $\ttthreeb$) are treated independently in the fit.  Furthermore, the overall acceptance and shape effects are estimated separately.  The impact of producing the additional \bjets by an NLO QCD calculation is evaluated using the $t\bar{t}b\bar{b}$ four-flavour scheme sample as listed in Table~\ref{tab:mcconfig}.
Additional uncertainties are evaluated from variations of the renormalisation and factorisation scales in the matrix element calculations by a factor of 0.5 and 2 and from the upward and downward variations of QCD scales for the initial- and final-state radiations as implemented in the parton shower model.  As the \ttbarj events are reweighted according to NNLO+NNLL QCD and NLO EW theory predictions in terms of top quark \pt and the \ttbar system \pt and mass, the corresponding uncertainties are applied in the analysis.  As mentioned in Section~\ref{sec:ttbar}, the normalisations of \ttb, \ttc and \ttl processes in each alternative \ttbar MC sample are scaled, before employing the corresponding distributions in the fit, to the respective estimated yields in the nominal \ttbar MC sample.

For the \Zjets process, which has a very small contribution in the signal regions, only the uncertainties in the \ZLF and \ZHF normalisation parameters are taken into account.  The \Wjets process gives a negligible contribution, so only a $5\%$ theory uncertainty in the production cross-section is taken into account~\cite{MATRIXv1}.   The multijet background has corresponding uncertainties assigned depending on the estimation method used for $1\tauh$ and $\geq\!2\tauh$ regions. The statistical and systematics uncertainties in the fake \tauh scale factors, as discussed in Section~\ref{sec:faketau}, are taken into account depending on the \tauh \pt and prongness. The diboson processes have negligible contributions in the signal and control regions. The uncertainties due to the modelling of these processes are evaluated by varying the QCD renormalisation and factorisation scales by a factor of 0.5 and 2, relative to the nominal scales. A $4\%$ uncertainty is taken in the inclusive production cross-section, and an additional uncertainty of $30\%$ is considered for the process with heavy-flavour jets~\cite{EXOT-2021-31}.

Theory uncertainties due to the modelling of \ttW, \ttZ, and \ttH processes are included in the analysis. The uncertainties are evaluated by comparing the fit distributions from nominal MC samples to those from the alternative MC generator predictions of each process, as listed in Table~\ref{tab:mcconfig}.
Further uncertainties are evaluated from the QCD renormalisation and factorisation scale variations by a factor of 0.5 and 2, relative to the nominal scales used in each MC sample.  Uncertainties of $50\%$~\cite{EXOT-2021-31}, $15\%$,~\cite{TOPQ-2018-08,TOPQ-2020-20} and $11\%$~\cite{deFlorian:2016spz} are assigned on the theory cross-sections for the \ttW, \ttZ, and \ttH processes, respectively.

Uncertainties in the modelling of other rare processes (such as $t (Z/\gamma^*)$, $t W (Z/\gamma^*)$, $t\bar t t$, $t\bar t W^+ W^-$, $VH$, triboson) are taken into account by assigning uncertainties up to $50\%$ in the production cross-sections. These are conservative uncertainties to account for limited information on the correct modelling of these background processes. The choice of the size of these uncertainties has a negligible impact on the results.  The $t\bar t t\bar t$ cross-section is scaled to the measured value~\cite{TOPQ-2021-08} and a $29\%$ uncertainty is assigned as normalisation uncertainty on this background.

\begin{table}[htbp!]
\begin{center}
\caption{Contribution of different uncertainties to the combined likelihood fit for the observed signal strength $\mu$, assuming a VLL mass of either $\SI{0.4}{\TeV}$ or $\SI{1}{\TeV}$. The impact of each nuisance parameter (NP) is computed by comparing the nominal best-fit value of $\mu$ with the result of the fit when fixing the considered NP to its best-fit value shifted by its post-fit uncertainty.
The impact of a group of NPs is obtained by summing in quadrature the impacts of all nuisance parameters in this category.  The fractional contribution of an uncertainty source refers to the ratio of the uncertainty in $\mu$ resulting from the fit when fixing the corresponding NP to the uncertainty in the best-fit value of $\mu$.  The ‘Other background’ category refer to the uncertainties from the fake \tauh and the rare SM background contributions such as $t\bar{t}W$ and $t\bar{t}t\bar{t}$, the relative contribution of these backgrounds vary with the signal mass.
}
\label{tab:systimpact}
\begin{tabular}{l|c|c}
\toprule
Uncertainty component & Fractional contribution [$\%$] & Fractional contribution [$\%$] \\
& \mVLL = 0.4~TeV & \mVLL = 1~TeV \\
\toprule
\textbf{Experimental} & & \\
Jet energy scale  & 5.2 & 3.5  \\
Jet energy resolution & 4.6 & 1.9  \\
\tauh energy scale & 9.7 & 1.1  \\
\tauh identification & 6.3 & 2.5  \\
Flavour tagging & 6.0 & 4.3  \\
BJET trigger scale factors & 6.2 & 2.1  \\
STT trigger scale factors & 0.2 & 0.3 \\
DTT trigger scale factors & 0.8 & 1.6  \\
\met scale and resolution & 1.7  & 0.8 \\
Others (JVT,  & 4.3  & 1.1 \\
pileup reweighting, luminosity) & & \\
\midrule
\textbf{Signal and background modelling} & & \\
Background normalisations  & 34 & 16 \\
\ttbarj modelling & 47 & 11  \\
QCD modelling & 14 & 8  \\
Other background & 10 & 11  \\
Signal modelling & 0.1 & 0.1  \\
\midrule
\textbf{MC statistics} & 34 & 39  \\
\midrule
\textbf{Total systematics} & 74 & 46  \\
\textbf{Data statistics} & 67 & 88  \\
\bottomrule
\end{tabular}
\end{center}
\end{table}

The impact of various categories of uncertainties on the observed signal-strength parameter $\mu$ is quantified.  The parameter $\mu$ is defined as a multiplicative factor applied to the predicted yield for the pair-produced VLL signal. Table~\ref{tab:systimpact} summarises the impact on $\mu$ for the $\SI{0.4}{\TeV}$ and $\SI{1}{\TeV}$ VLL signal hypotheses.  The impact of each NP on the $\mu$ is evaluated by comparing the nominal best-fit value of $\mu$ with the result of the fit when fixing the considered NP to its best-fit value shifted by its post-fit uncertainty.


%

%
\section{Results}
\label{sec:result}

A maximum-likelihood fit is performed for each signal hypothesis on all bins of the seven CRs and five SRs defined in Section~\ref{sec:analysis} to determine the VLL signal cross-section and the normalisation factors of the \ttb, \ttc, \ttl, \ZLF, \ZHF, Multijet 1$\tauh \geq\!3b$ MST, Multijet $1\tauh \geq\!3b$ BJET and Multijet $2\tauh\geq\!3b$ MSDT processes. The Multijet $2\tauh2b$ control region is binned in the \sumPCT distribution, whereas the total event yield (i.e. a single bin) is used in the remaining control regions.

The likelihood function ${\cal L}(\mu,\vec{\lambda}, \vec{\theta})$ is constructed as a product of Poisson probability terms over all bins considered in the search.  It depends on  $\mu$, $\vec{\lambda}$, the normalisation factors for several backgrounds (see Section~\ref{sec:backgrounds}), and $\vec{\theta}$, a set of NPs encoding systematic uncertainties in the signal and background expectations~\cite{Cranmer:2012sba}. The predicted signal yield depends on the assumed mass \mVLL. Systematic uncertainties, summarised in Section~\ref{sec:systematics}, can impact the estimated signal and background rates, the migration of events between categories, and the shape of the fitted distributions.
Both $\mu$ and $\vec{\lambda}$ are treated as free parameters in the likelihood fit. The NPs $\vec{\theta}$ allow variations of the expectations for signal and background according to the systematic uncertainties, subject to Gaussian constraints in the likelihood fit.
Their fitted values represent the deviations from the nominal expectations that globally provide the best fit to the data.
Statistical uncertainties in each bin due to the limited size of the simulated samples are taken into account by dedicated parameters using the Beeston--Barlow `lite' technique~\cite{Barlow:1993dm}.

The test statistic $q_\mu$ is defined as the profile likelihood ratio:
$q_\mu = -2\ln({\cal L}(\mu,{\hat{\vec{\lambda}}}_\mu, {\hat{\vec{\theta}}}_\mu)/{\cal L}(\hat{\mu},\hat{\vec{\lambda}}_{\hat{\mu}}, \hat{\vec{\theta}}_{\hat{\mu}}))$,
where $\hat{\mu}$, $\hat{\vec{\lambda}}_{\hat{\mu}}$, and $\hat{\vec{\theta}}_{\hat{\mu}}$ are the values of the parameters that maximise the likelihood function,
and ${\hat{\vec{\lambda}}}_\mu$ and ${\hat{\vec{\theta}}}_\mu$ are the values of the parameters that maximise the likelihood function for a given value of $\mu$.
The test statistic $q_\mu$ is evaluated with the {\textsc RooFit} package~\cite{RooFit}.
A related statistic is used to determine the probability that the observed data are compatible with the background-only hypothesis (i.e.~the discovery test) by setting $\mu=0$ in the profile likelihood ratio ($q_0$).
The $p$-value (referred to as $p_0$) representing the probability of the data being compatible with the background-only hypothesis is approximated using the asymptotic formulae given in Ref.~\cite{Cowan:2010js}, above the observed value of $q_0$.

\begin{figure}
\centering
\subfloat[]{\includegraphics[width=.45\textwidth]{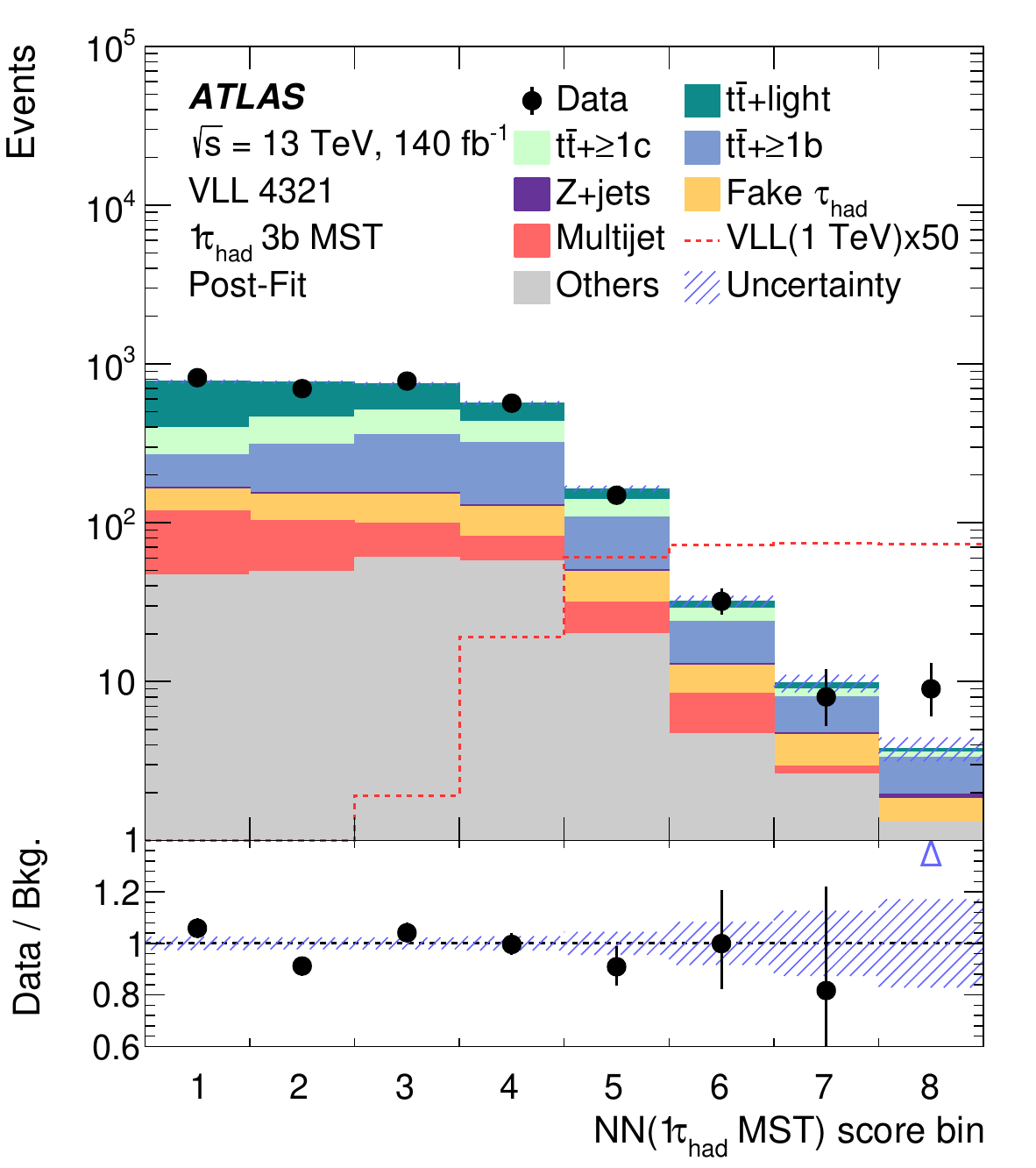}}
\subfloat[]{\includegraphics[width=.45\textwidth]{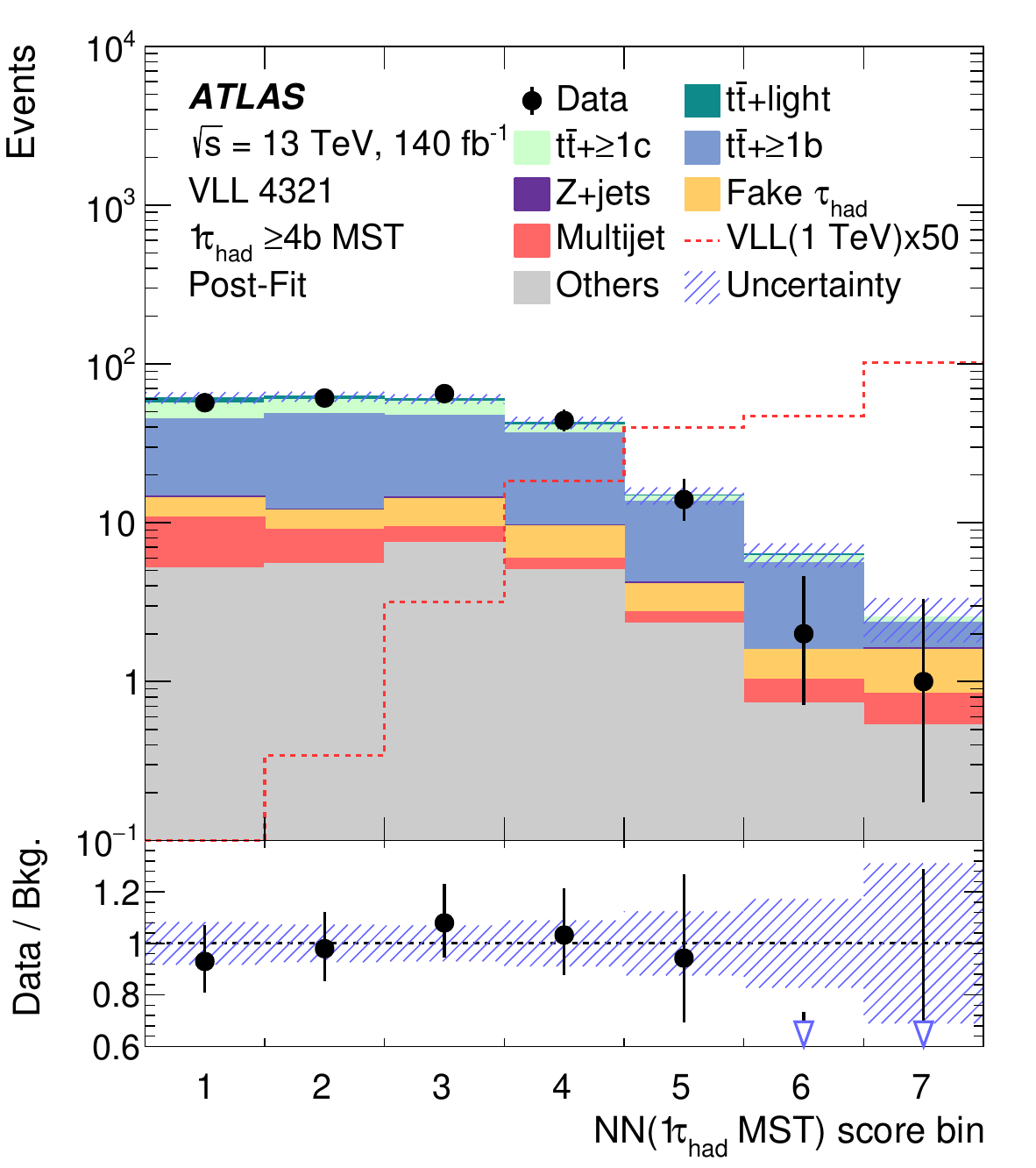}} \\
\subfloat[]{\includegraphics[width=.45\textwidth]{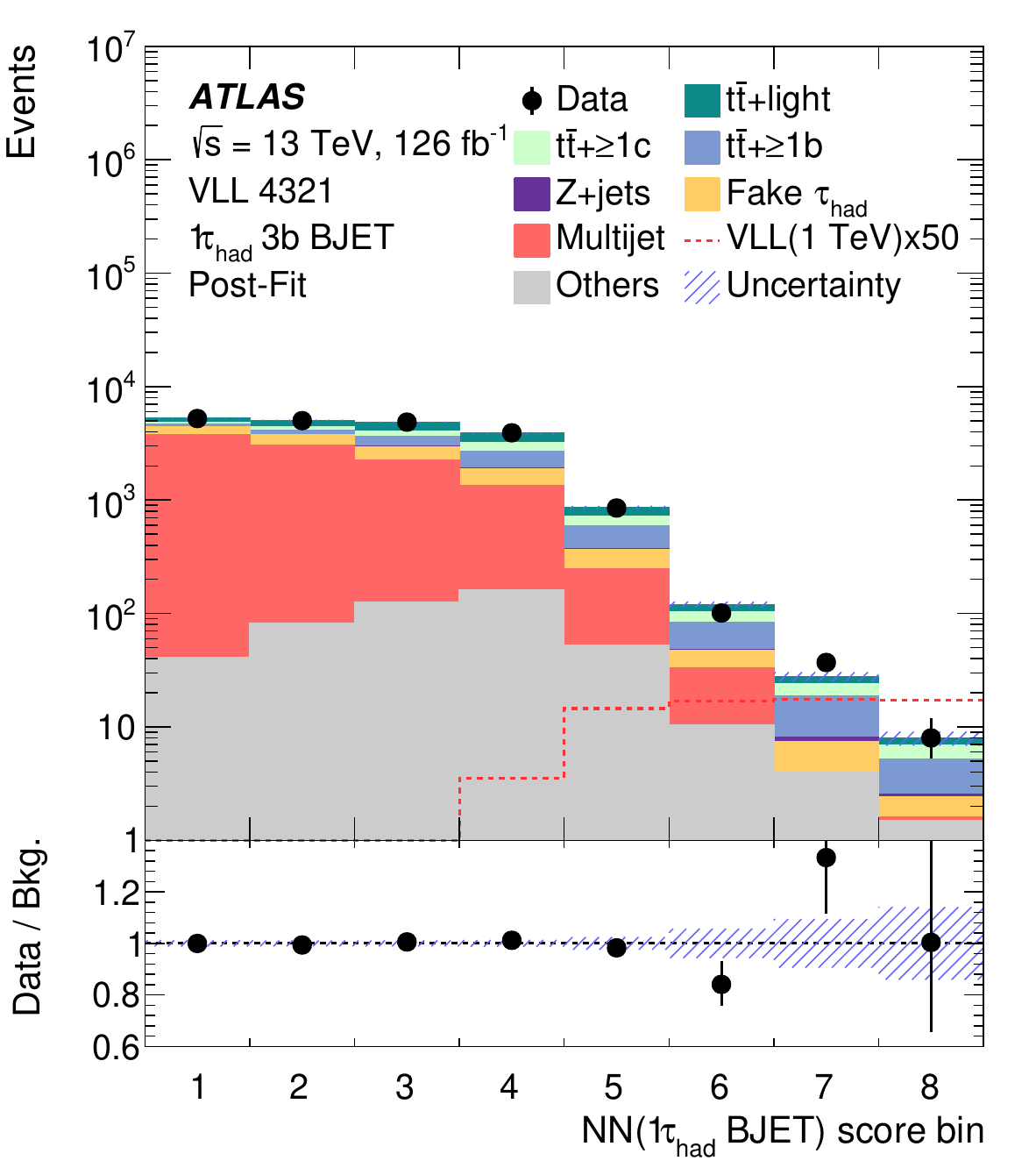}}
\subfloat[]{\includegraphics[width=.45\textwidth]{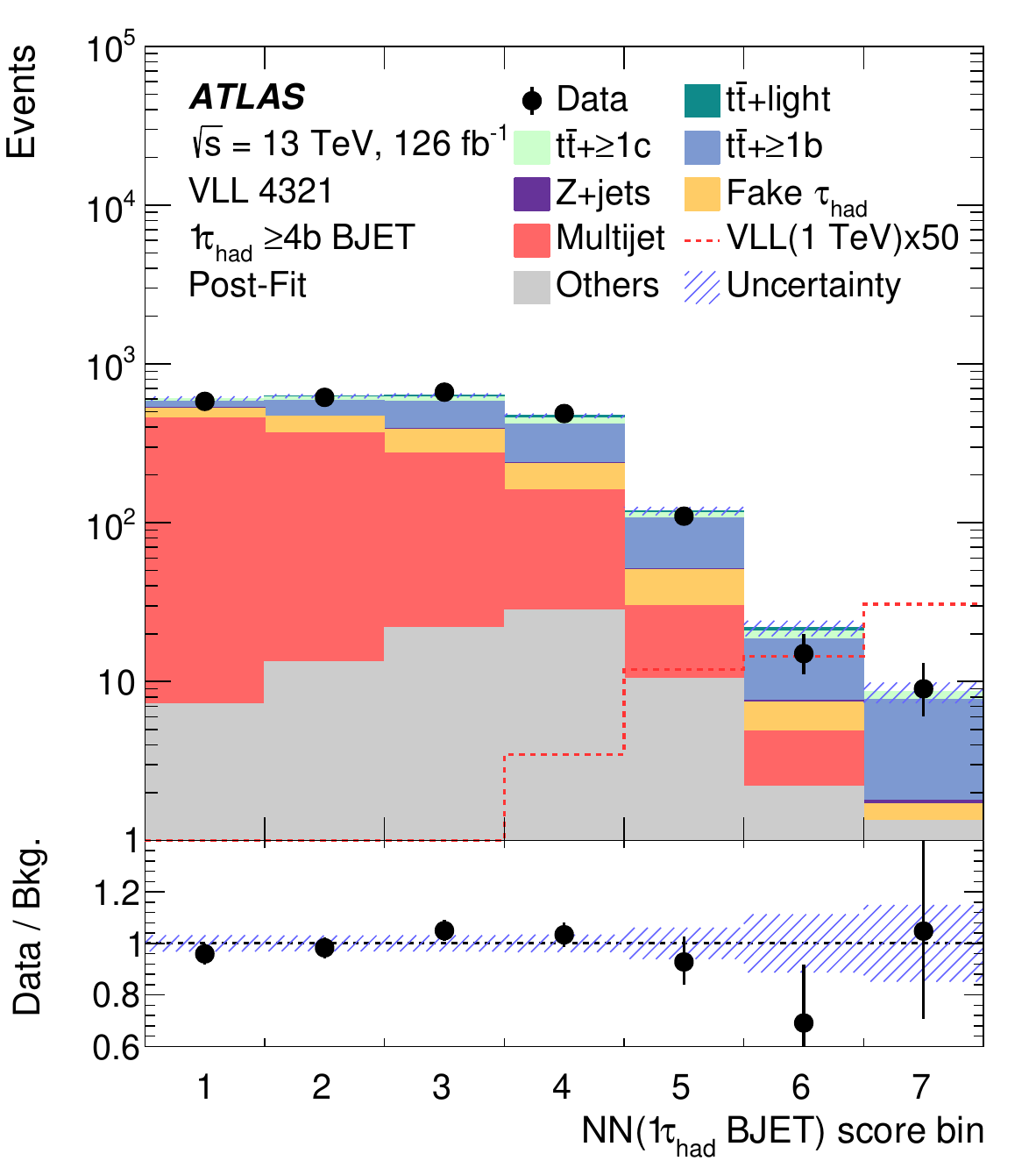}}
\caption{Comparison between data and the background prediction for the event yields in all four $1\tauh$ signal region categories: (a) $1\tauh 3b$ MST, (b) $1\tauh \geq\!4b$ MST, (c) $1\tauh 3b$ BJET, and (d) $1\tauh \geq\!4b$ BJET, with the NN distributions corresponding to \mVLL=1~TeV. The background contributions after the likelihood fit to data (`Post-Fit') under the background-only hypothesis are shown as filled histograms. The ratio of the data to the background (`Bkg.') prediction is shown in the lower panel. The size of the combined statistical and systematic uncertainty in the background prediction is indicated by the blue hatched band. The blue triangles indicate points that are outside the vertical range of the figure.}
\label{fig:signal_regions_1tau}
\end{figure}

\begin{figure}
\centering
\includegraphics[width=.45\textwidth]{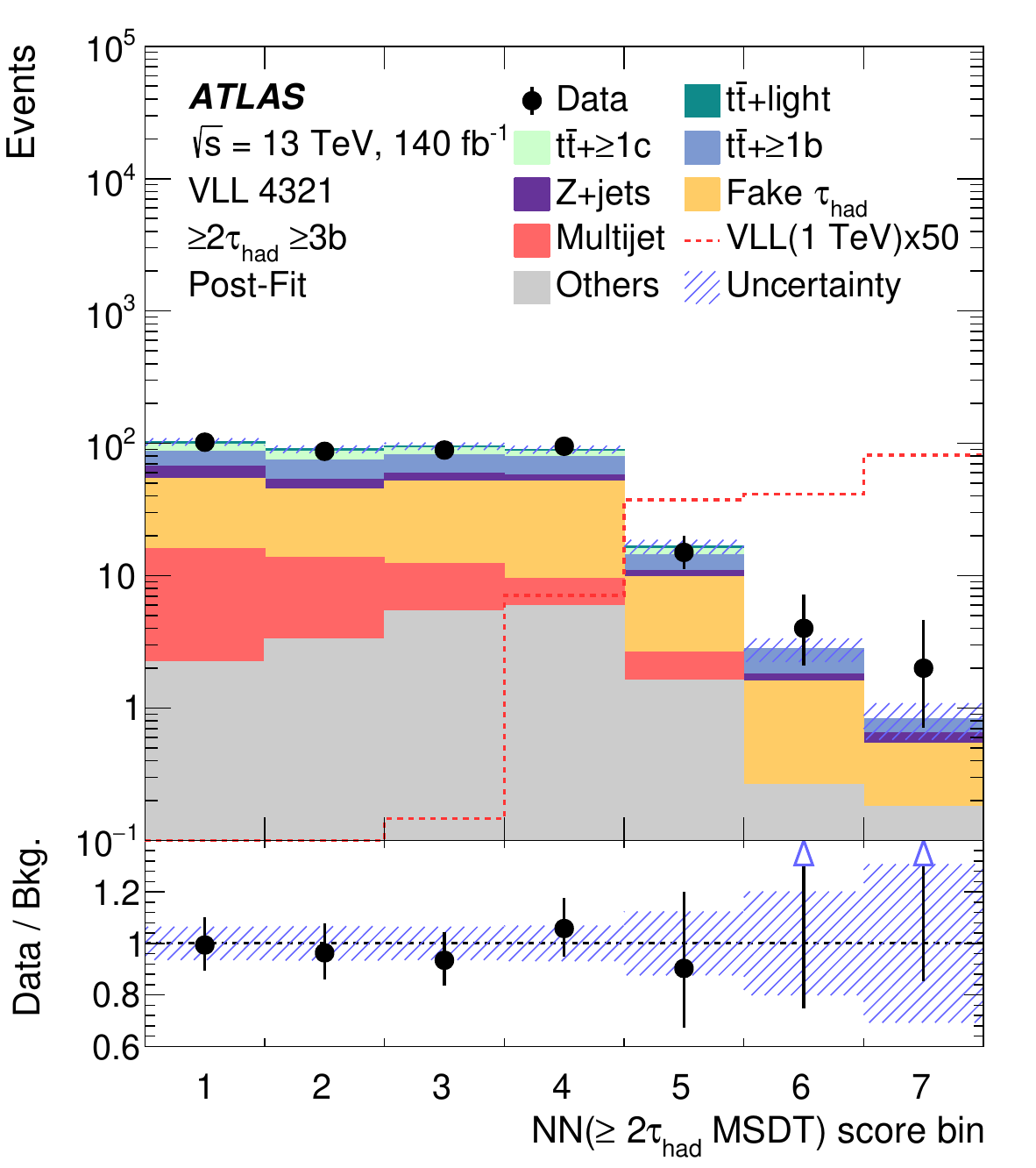}
\caption{Comparison between data and the background prediction for the event yields in the $2\tauh \geq\!3b$ MSDT signal region category with the NN distribution corresponding to \mVLL=1~TeV. The background contributions after the likelihood fit to data (`Post-Fit') under the background-only hypothesis are shown as filled histograms. The ratio of the data to the background (`Bkg.') prediction is shown in the lower panel. The size of the combined statistical and systematic uncertainty in the background prediction is indicated by the blue hatched band. The blue triangles indicate points that are outside the vertical range of the figure.}
\label{fig:signal_regions_2tau}
\end{figure}

\begin{figure*}[tbp]
\centering
\includegraphics[width=0.6\textwidth]{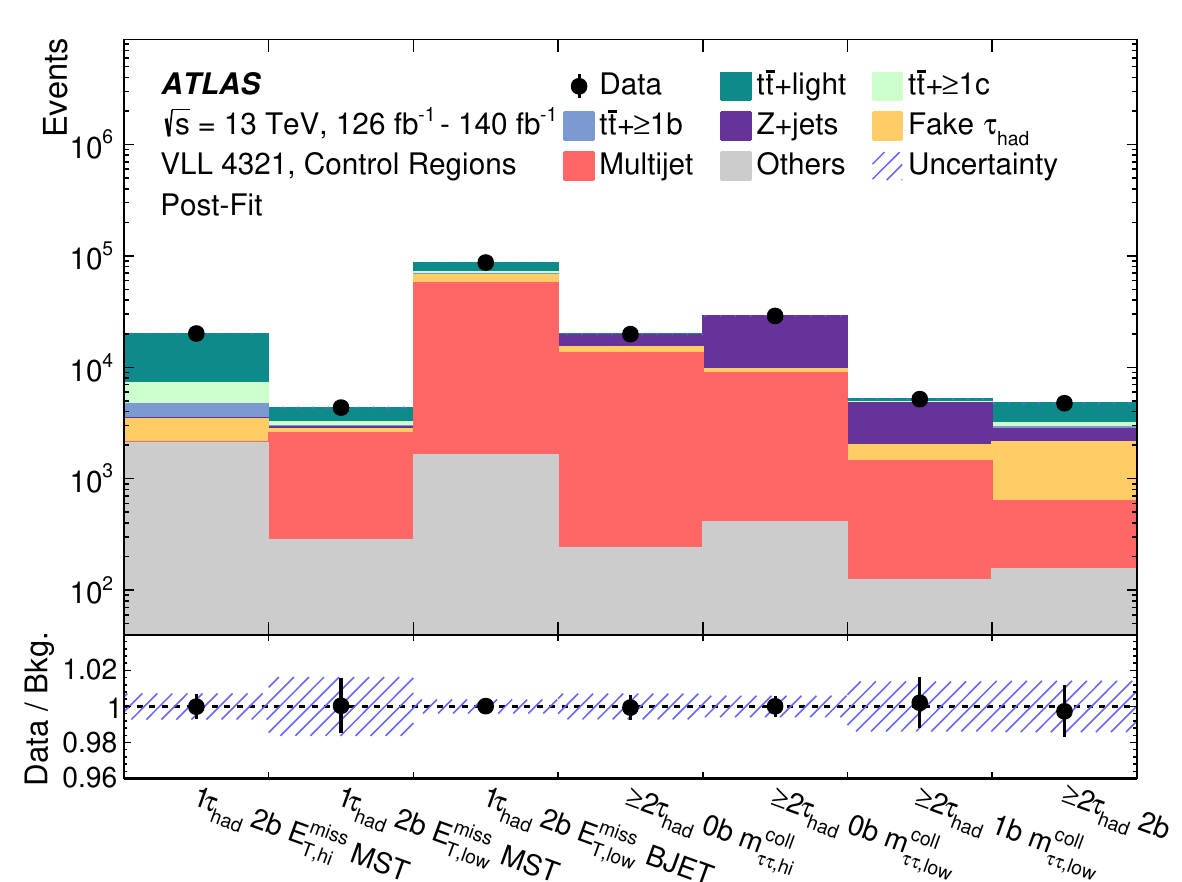} \\
\caption{Comparison between data and the background prediction for the event yields in the seven control region categories.
The ratio of the data to the background (`Bkg.') prediction is shown in the lower panel. %
The size of the combined statistical and systematic uncertainty in the background prediction is indicated by the blue hatched band.
}
\label{fig:summarySRCR}
\end{figure*}

\begin{table}[tbp]
\caption{Summary of observed and predicted yields in the five signal region categories, corresponding to their respective data luminosities. The background prediction is shown after the combined likelihood fit to data under the background-only hypothesis across all control region and signal region categories, while the NN score is evaluated for a $\SI{1}{\TeV}$ mass in the signal region. The expected signal yields that are obtained by using their theory cross-sections are also shown with their pre-fit uncertainties, assuming $\mu$=1 for an \mVLL mass of $\SI{1}{\TeV}$. The uncertainties correspond to the combined statistical and systematic uncertainties in the predicted yields. Total background yield uncertainties are smaller than the quadratic sum of individual backgrounds due to correlations in uncertainty components and normalisation factors.  The `Others' contribution is dominated by \tttt and \ttW. \label{tab:SRspostfityields}}
\begin{center}
\resizebox{1\linewidth}{!}{%
\begin{tabular}{l
S[table-format=4.2, table-number-alignment=right]@{$\,\pm\,$}S[table-format=3.1]
S[table-format=3.2, table-number-alignment=right]@{$\,\pm\,$}S[table-format=2.1]
S[table-format=5.2, table-number-alignment=right]@{$\,\pm\,$}S[table-format=3.2]
S[table-format=3.2, table-number-alignment=right]@{$\,\pm\,$}S[table-format=3.2]
S[table-format=3.2, table-number-alignment=right]@{$\,\pm\,$}S[table-format=3.1]
}
\hline
\hline
& \multicolumn{2}{c}{1$\tauh 3b$ MST}  & \multicolumn{2}{c}{1$\tauh \geq\!4b$ MST}  & \multicolumn{2}{c}{1$\tauh 3b$ BJET} & \multicolumn{2}{c}{1$\tauh \geq\!4b$ BJET} & \multicolumn{2}{c}{$\geq\!2\tauh \geq\!3b$ MSDT}  \\
& \multicolumn{2}{c}{(140 fb$^{-1}$)} & \multicolumn{2}{c}{(140 fb$^{-1}$)} & \multicolumn{2}{c}{(126 fb$^{-1}$)} & \multicolumn{2}{c}{(126 fb$^{-1}$)} & \multicolumn{2}{c}{(140 fb$^{-1}$)} \\
\hline
\hline
Data   & \multicolumn{2}{c}{3062} & \multicolumn{2}{c}{244} & \multicolumn{2}{c}{20020} & \multicolumn{2}{c}{2482} & \multicolumn{2}{c}{394} \\
\hline
Total background &  3070 & 50 & 250 & 13 & 20010 & 140 & 2490 & 50 & 398 & 18 \\
\hline
\ttb &  720 & 140 & 141 & 22 & 2300 & 500 & 620 & 90 & 90 & 18 \\
\ttc &  590 & 230 & 41 & 17 & 1600 & 600 & 130 & 50 & 41 & 17 \\
\ttl &  1080 & 120 & 10.3 & 2.9 & 2650 & 340 & 45 & 11 & 15.5 & 3.3 \\
\Zjets &  16.6 & 1.7 & 1.08 & 0.16 & 79 & 14 & 8.4 & 1.8 & 36.0 & 3.1  \\
Multijets &  200 & 40 & 13 & 5 & 10200 & 500 & 1210 & 70 & 36 & 9 \\
Others &  240 & 50 & 27 & 4 & 480 & 110 & 85 & 12 & 18.9 & 3.1 \\
Fake \tauh &  214 & 23 & 18 & 4 & 2730 & 270 & 390 & 40 & 160 & 18 \\
\hline
\hline
$\mVLL$ 1.0 TeV  &  6.07 & 0.23 & 4.37 & 0.24 & 1.32 & 0.13 & 1.10 & 0.14 & 3.41 & 0.19  \\
\hline
\hline
\end{tabular}
}
\end{center}
\end{table}

A comparison of the distributions of observed and expected events in the five SRs is presented in Figures~\ref{fig:signal_regions_1tau} and~\ref{fig:signal_regions_2tau} after the combined likelihood fit under the background-only hypothesis, considering a mass of 1~TeV in the evaluation of NN score distribution. Overall, there is good agreement between the data and the predicted background yields across all event categories, with the lowest $p$-value for data-background compatibility of 0.043 observed in the $1\tauh 3b$ MST region. %
After the fit, no NPs show significant deviations or constraints from their nominal values, indicating that all the systematic uncertainties considered in the analysis are well-behaved.  There are also no strong correlation factors among the fitted values of the normalisation parameters of \ttbarj, \Zjets and multijet background. The main uncertainty in this analysis comes from the limited data statistics.
The corresponding post-fit yields for the five SRs can be found in Table~\ref{tab:SRspostfityields}.  The data and expected yields in the seven CRs after the combined likelihood fit under the background-only hypothesis are shown in Figure~\ref{fig:summarySRCR}.  As summarised in Table~\ref{tab:systimpact}, the systematic uncertainty with the largest impact on the signal strength originates from the background MC statistical uncertainty as well as from the \ttbarj background modelling. %
In general, good agreement between the data and predicted background yields is found across all event categories. The observed $p_0$ is checked for each explored signal scenario, and the smallest value is found to be 0.13, corresponding to a local significance of $1.1\sigma$, for the 400~GeV signal point.

\begin{figure*}[t]
\centering
\includegraphics[width=0.7\textwidth]{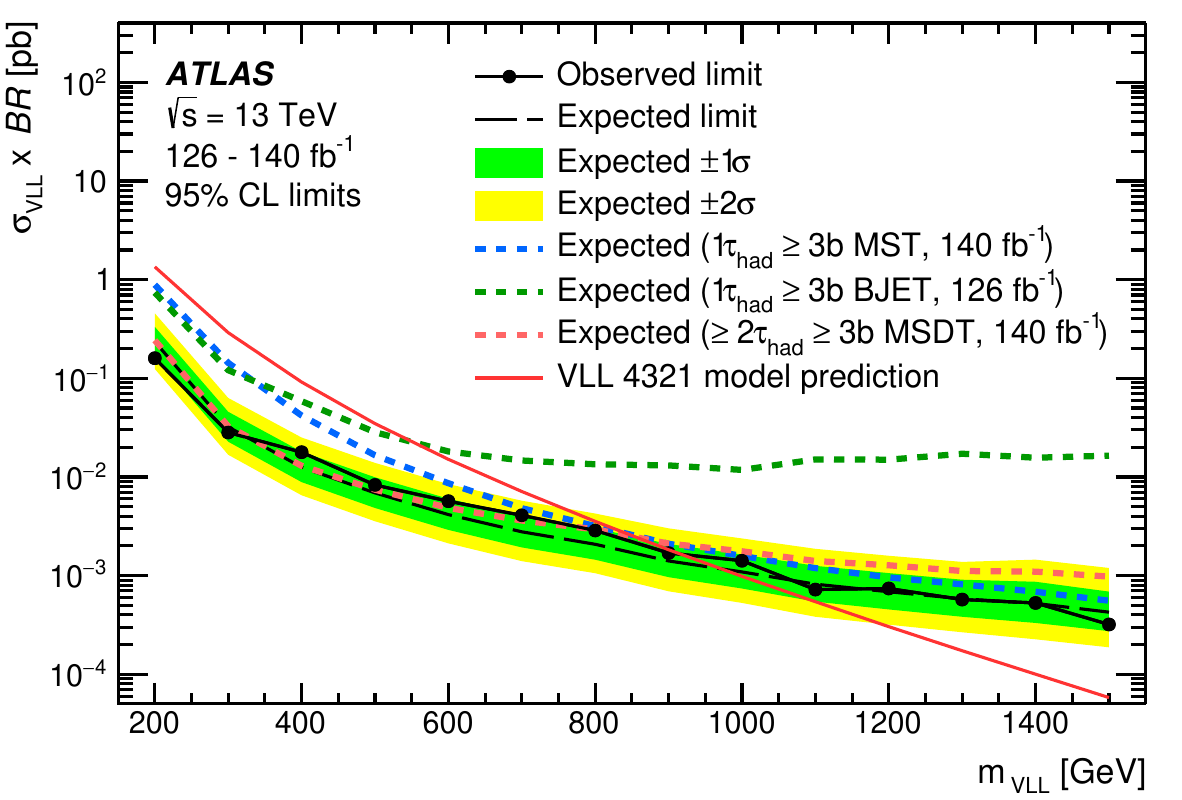}
\caption{
Observed (solid line with markers) and expected (dashed line) 95\% CL upper limits on the VLL pair production cross-section ($\sigma_{\mathrm{VLL}}$) times branching ratio ($BR$) to third generation quarks and leptons as a function of \mVLL. The limits presented in black lines are obtained after combining all five signal regions.  The inner green (outer yellow) band corresponds to the $\pm$1\,$\sigma$ ($\pm$2\,$\sigma$) uncertainty around the combined expected limit. The 95\% CL expected upper limits in the three individual channels ($1\tauh \geq\!3b$ MST, $1\tauh \geq\!3b$ BJET and $\geq\!2\tauh \geq\!3b$ MSDT) are shown for comparison.  The solid red line represents the theory prediction of the VLL pair production cross-section at NLO in QCD.%
}
\label{fig:limits}
\end{figure*}

In the absence of any significant excess, upper limits on the signal production cross-section for each of the signal scenarios considered are derived by using $q_\mu$ in the CL$_{\textrm{s}}$ method~\cite{Junk:1999kv,Read:2002hq}.
For a given signal scenario, values of the production cross-section (parameterised by $\mu$) yielding CL$_{\textrm{s}} < 0.05$, where CL$_{\textrm{s}}$ is computed using the asymptotic approximation~\cite{Cowan:2010js},
are excluded at $\geq\!95\%$ confidence level (CL).

Figure~\ref{fig:limits} shows the 95\% CL upper limits on the VLL pair production cross-section times branching ratio as a function of \mVLL. The observed and expected 95\% CL lower limits on \mVLL are $\SI{910}{\GeV}$ and $\SI{970}{\GeV}$, respectively.
When accounting for systematic uncertainties, the cross-section limits degrade by only $8\%$ for VLL masses around $\SI{1}{\TeV}$, and they degrade by $25\%--70\%$ for VLL masses below $\SI{0.6}{\TeV}$.  The upper limits on the VLL production cross-section, calculated using pseudo-experiments for the CL$_{\textrm{s}}$ method, are consistent with the presented results within $5\%$ in the most sensitive regions. However, for \mVLL below 600 GeV or above 1100 GeV, the limits derived from pseudo-experiments are up to $20\%$ higher due to low background yields in the most sensitive bins.

This analysis, the first search from the ATLAS Collaboration for VLLs decaying through $U_1$ to third generation SM fermions, excludes the $\SI{600}{\GeV}$ VLL mass, which is a representative point within the mass range where the CMS Collaboration reports an excess, including a $2.8 \sigma$ excess at this specific mass~\cite{CMS-B2G-21-004}.
The observed $p_0$ in the analysis presented here for the same representative signal mass point is 0.18, corresponding to a local significance of $0.9 \sigma$. The $95\%$ CL observed (expected) upper limit on the signal cross-section for the VLL mass of $\SI{600}{\GeV}$ is 5.8 fb (4.1 fb). In the CMS Collaboration analysis~\cite{CMS-B2G-21-004}, the corresponding expected upper limit is about 10 fb, with the most sensitive channel selected with: a trigger requiring three $b$-tagged jets and event selections with at least four jets (\pt thresholds of 80, 65, 50, and 50~GeV), three or more of which are $b$-tagged jets, along with two or more \tauh (\pt $> \SI{20}{\GeV}$), $H_{\mathrm{T,jets}} > \SI{400}{\GeV}$ and $\met > \SI{40}{\GeV}$.  The ATLAS search sensitivity is driven by the $\geq\!2\tauh \geq\!3b$ MSDT channel in the investigated mass range ($\SI{200}{\GeV}$--$\SI{1500}{\GeV}$). At high mass (\mVLL $>$ $\SI{700}{\GeV}$), the $1 \tauh \geq\!3b$ MST channels reinforce the sensitivity further as illustrated in Figure~\ref{fig:limits}.

\begin{figure*}[t]
\centering
\includegraphics[width=0.7\textwidth]{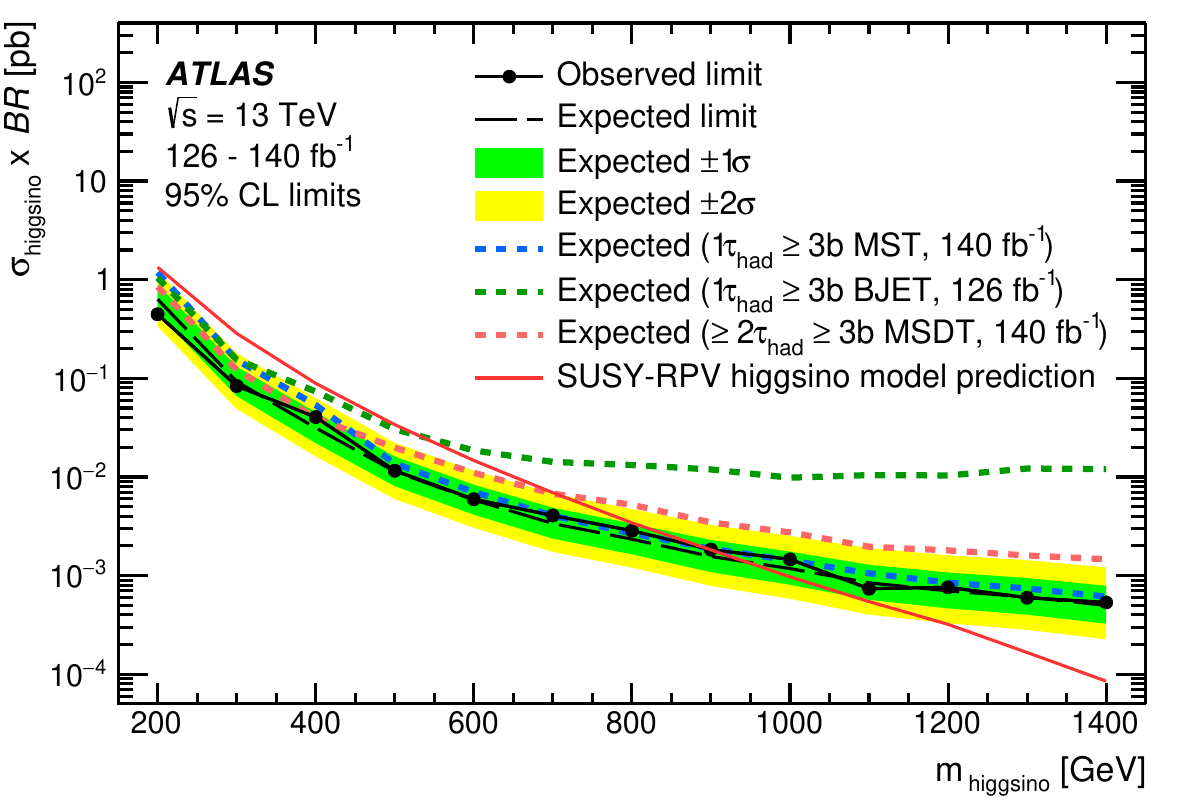}
\caption{
Observed (solid line with markers) and expected (dashed line) 95\% CL upper limits on the higgsino pair production cross-section ($\sigma_{\mathrm{higgsino}}$) times branching ratio ($BR$) to third generation quarks and leptons as a function of $m_{\mathrm{higgsino}}$. The limits presented in black lines are obtained after combining all five signal regions.  The inner green (outer yellow) band corresponds to the $\pm$1\,$\sigma$ ($\pm$2\,$\sigma$) uncertainty around the combined expected limit. The 95\% CL  expected upper limits in the three individual channels ($1\tauh \geq\!3b$ MST, $1\tauh \geq\!3b$ BJET and $\geq\!2\tauh \geq\!3b$ MSDT) are shown for comparison.  The solid red line represents the theory prediction of the higgsino pair production cross-section at NLO in QCD.
}
\label{fig:limits_higgsino}
\end{figure*}

\begin{figure*}[t]
\centering
\includegraphics[width=0.7\textwidth]{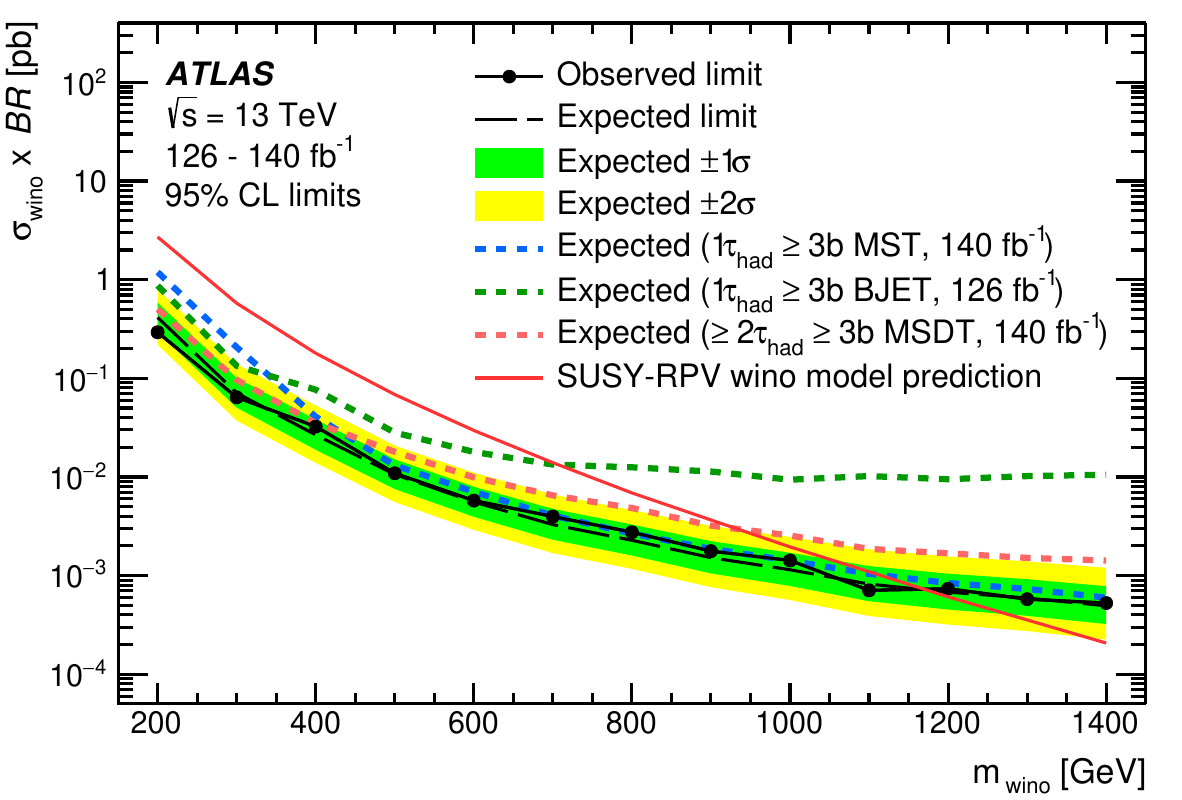}
\caption{
Observed (solid line with markers) and expected (dashed line) 95\% CL upper limits on the wino pair production cross-section ($\sigma_{\mathrm{wino}}$) times branching ratio ($BR$) to third generation quarks and leptons as a function of $m_{\mathrm{wino}}$. The limits presented in black lines are obtained after combining all five signal regions.  The surrounding inner green (outer yellow) band corresponds to the $\pm$1\,$\sigma$ ($\pm$2\,$\sigma$) uncertainty around the combined expected limit. The 95\% CL expected upper limits in the three individual channels ($1\tauh \geq\!3b$ MST, $1\tauh \geq\!3b$ BJET and $\geq\!2\tauh \geq\!3b$ MSDT) are shown for comparison.  The solid red line represents the theory prediction of the wino pair production cross-section at NLO in QCD.
}
\label{fig:limits_wino}
\end{figure*}

The NN optimised for the VLL signal is also applied to the RPV SUSY signal, with the corresponding distributions from the signal and control regions included in the simultaneous fit. The resulting cross-section times branching ratio limits are derived for electroweakino pair production under the higgsino or wino mixing assumptions in the RPV SUSY models, as shown in Figures~\ref{fig:limits_higgsino} and~\ref{fig:limits_wino}.
The kinematic distributions of the final states from these SUSY models, where the chargino/neutralino decays into third generation fermions, are close to those obtained from the VLL pair processes. The signal cross-sections in these models show only minimal differences compared to the background size, and the background normalisation factors obtained from the fit remain unchanged.
The $95\%$ CL lower limits were set on the higgsino mass $m_{\mathrm{higgsino}}$ of 880 GeV (940 GeV) and on the wino mass $m_{\mathrm{wino}}$ of 1170 GeV (1170 GeV) for the observed (expected) values.


%

%

%
%
%

%
%
\section{Conclusion}
\label{sec:conclusion}

A search for electroweak production of vector-like leptons as predicted in the \vll model is performed using the $\sqrt{s}=\SI{13}{\TeV}$ dataset collected with the ATLAS detector at the LHC, corresponding to an integrated luminosity of \lumivalbjet to \lumival. The signal regions characterised by the presence of \bjets and hadronically decaying $\tau$-leptons are optimised using neural networks parameterised in the generated mass of the vector-like lepton signal.  The likelihood fits to data using the discriminant score distributions across the signal and control regions show good consistency of data with the SM background predictions. Upper limits at $95\%$ CL on the production cross-section times branching ratio into third generation SM fermions are derived as a function of vector-like lepton mass, setting a lower observed (expected) limit on the vector-like lepton mass at $\SI{910}{\GeV}$ ($\SI{970}{\GeV}$).  The results are also interpreted in context of RPV SUSY models leading to similar final states,
for which the observed (expected) lower limits on the higgsino mass in the higgsino mixing model is 880 GeV (940 GeV) and on the wino mass in the wino mixing model is 1170 GeV (1170 GeV).


%

%

%
%
%
%
%
\section*{Acknowledgements}
%

%

%
%

%
%

We thank CERN for the very successful operation of the LHC and its injectors, as well as the support staff at
CERN and at our institutions worldwide without whom ATLAS could not be operated efficiently.

The crucial computing support from all WLCG partners is acknowledged gratefully, in particular from CERN, the ATLAS Tier-1 facilities at TRIUMF/SFU (Canada), NDGF (Denmark, Norway, Sweden), CC-IN2P3 (France), KIT/GridKA (Germany), INFN-CNAF (Italy), NL-T1 (Netherlands), PIC (Spain), RAL (UK) and BNL (USA), the Tier-2 facilities worldwide and large non-WLCG resource providers. Major contributors of computing resources are listed in Ref.~\cite{ATL-SOFT-PUB-2025-001}.

We gratefully acknowledge the support of ANPCyT, Argentina; YerPhI, Armenia; ARC, Australia; BMWFW and FWF, Austria; ANAS, Azerbaijan; CNPq and FAPESP, Brazil; NSERC, NRC and CFI, Canada; CERN; ANID, Chile; CAS, MOST and NSFC, China; Minciencias, Colombia; MEYS CR, Czech Republic; DNRF and DNSRC, Denmark; IN2P3-CNRS and CEA-DRF/IRFU, France; SRNSFG, Georgia; BMFTR, HGF and MPG, Germany; GSRI, Greece; RGC and Hong Kong SAR, China; ICHEP and Academy of Sciences and Humanities, Israel; INFN, Italy; MEXT and JSPS, Japan; CNRST, Morocco; NWO, Netherlands; RCN, Norway; MNiSW, Poland; FCT, Portugal; MNE/IFA, Romania; MSTDI, Serbia; MSSR, Slovakia; ARIS and MVZI, Slovenia; DSI/NRF, South Africa; MICIU/AEI, Spain; SRC and Wallenberg Foundation, Sweden; SERI, SNSF and Cantons of Bern and Geneva, Switzerland; NSTC, Taipei; TENMAK, T\"urkiye; STFC/UKRI, United Kingdom; DOE and NSF, United States of America.

Individual groups and members have received support from BCKDF, CANARIE, CRC and DRAC, Canada; CERN-CZ, FORTE and PRIMUS, Czech Republic; COST, ERC, ERDF, Horizon 2020, ICSC-NextGenerationEU and Marie Sk{\l}odowska-Curie Actions, European Union; Investissements d'Avenir Labex, Investissements d'Avenir Idex and ANR, France; DFG and AvH Foundation, Germany; Herakleitos, Thales and Aristeia programmes co-financed by EU-ESF and the Greek NSRF, Greece; BSF-NSF and MINERVA, Israel; NCN and NAWA, Poland; La Caixa Banking Foundation, CERCA Programme Generalitat de Catalunya and PROMETEO and GenT Programmes Generalitat Valenciana, Spain; G\"{o}ran Gustafssons Stiftelse, Sweden; The Royal Society and Leverhulme Trust, United Kingdom.

In addition, individual members wish to acknowledge support from Armenia: Yerevan Physics Institute (FAPERJ); CERN: European Organization for Nuclear Research (CERN DOCT); Chile: Agencia Nacional de Investigaci\'on y Desarrollo (FONDECYT 1230812, FONDECYT 1240864); China: Chinese Ministry of Science and Technology (MOST-2023YFA1605700, MOST-2023YFA1609300), National Natural Science Foundation of China (NSFC - 12175119, NSFC 12275265); Czech Republic: Czech Science Foundation (GACR - 24-11373S), Ministry of Education Youth and Sports (ERC-CZ-LL2327, FORTE CZ.02.01.01/00/22\_008/0004632), PRIMUS Research Programme (PRIMUS/21/SCI/017); EU: H2020 European Research Council (ERC - 101002463); European Union: European Research Council (BARD No. 101116429, ERC - 948254, ERC 101089007), European Regional Development Fund (SMASH COFUND 101081355, SLO ERDF), Horizon 2020 Framework Programme (MUCCA - CHIST-ERA-19-XAI-00), European Union, Future Artificial Intelligence Research (FAIR-NextGenerationEU PE00000013), Horizon 2020 (EuroHPC - EHPC-DEV-2024D11-051), Italian Center for High Performance Computing, Big Data and Quantum Computing (ICSC, NextGenerationEU); France: Agence Nationale de la Recherche (ANR-21-CE31-0022, ANR-22-EDIR-0002); Germany: Baden-Württemberg Stiftung (BW Stiftung-Postdoc Eliteprogramme), Deutsche Forschungsgemeinschaft (DFG - 469666862, DFG - CR 312/5-2); China: Research Grants Council (GRF); Italy: Istituto Nazionale di Fisica Nucleare (ICSC, NextGenerationEU), Ministero dell'Università e della Ricerca (NextGenEU 153D23001490006 M4C2.1.1, NextGenEU I53D23000820006 M4C2.1.1, NextGenEU I53D23001490006 M4C2.1.1); Japan: Japan Society for the Promotion of Science (JSPS KAKENHI JP22H01227, JSPS KAKENHI JP22H04944, JSPS KAKENHI JP22KK0227, JSPS KAKENHI JP23KK0245, JSPS KAKENHI JP24K23939); Norway: Research Council of Norway (RCN-314472); Poland: Ministry of Science and Higher Education (IDUB AGH, POB8, D4 no 9722), Polish National Science Centre (NCN 2021/42/E/ST2/00350, NCN OPUS 2023/51/B/ST2/02507, NCN OPUS nr 2022/47/B/ST2/03059, NCN UMO-2019/34/E/ST2/00393, UMO-2022/47/O/ST2/00148, UMO-2023/49/B/ST2/04085, UMO-2023/51/B/ST2/00920, UMO-2024/53/N/ST2/00869); Portugal: Foundation for Science and Technology (FCT); Spain: Generalitat Valenciana (Artemisa, FEDER, IDIFEDER/2018/048), Ministry of Science and Innovation (MCIN \& NextGenEU PCI2022-135018-2, MICIN \& FEDER PID2021-125273NB, RYC2019-028510-I, RYC2020-030254-I, RYC2021-031273-I, RYC2022-038164-I); Sweden: Carl Trygger Foundation (Carl Trygger Foundation CTS 22:2312), Swedish Research Council (Swedish Research Council 2023-04654, VR 2021-03651, VR 2022-03845, VR 2022-04683, VR 2023-03403, VR 2024-05451), Knut and Alice Wallenberg Foundation (KAW 2018.0458, KAW 2022.0358, KAW 2023.0366); Switzerland: Swiss National Science Foundation (SNSF - PCEFP2\_194658); United Kingdom: Leverhulme Trust (Leverhulme Trust RPG-2020-004), Royal Society (NIF-R1-231091); United States of America: U.S. Department of Energy (ECA DE-AC02-76SF00515), Neubauer Family Foundation.

%
%


%
%
%
%
%
%
%
%
%

%
%

%
%
%
\printbibliography
\clearpage
 
\begin{flushleft}
\hypersetup{urlcolor=black}
{\Large The ATLAS Collaboration}

\bigskip

\AtlasOrcid[0000-0002-6665-4934]{G.~Aad}$^\textrm{\scriptsize 104}$,
\AtlasOrcid[0000-0001-7616-1554]{E.~Aakvaag}$^\textrm{\scriptsize 17}$,
\AtlasOrcid[0000-0002-5888-2734]{B.~Abbott}$^\textrm{\scriptsize 123}$,
\AtlasOrcid[0000-0002-0287-5869]{S.~Abdelhameed}$^\textrm{\scriptsize 119a}$,
\AtlasOrcid[0000-0002-1002-1652]{K.~Abeling}$^\textrm{\scriptsize 55}$,
\AtlasOrcid[0000-0001-5763-2760]{N.J.~Abicht}$^\textrm{\scriptsize 49}$,
\AtlasOrcid[0000-0002-8496-9294]{S.H.~Abidi}$^\textrm{\scriptsize 30}$,
\AtlasOrcid[0009-0003-6578-220X]{M.~Aboelela}$^\textrm{\scriptsize 45}$,
\AtlasOrcid[0000-0002-9987-2292]{A.~Aboulhorma}$^\textrm{\scriptsize 36e}$,
\AtlasOrcid[0000-0001-5329-6640]{H.~Abramowicz}$^\textrm{\scriptsize 157}$
% ...
\bigskip
\\

% ...
$^{17}$Department for Physics and Technology, University of Bergen, Bergen; Norway.\\
% ...
$^{30}$Physics Department, Brookhaven National Laboratory, Upton NY; United States of America.\\
% ...
$^{36}$$^{(a)}$Facult\'e des Sciences Ain Chock, Universit\'e Hassan II de Casablanca;$^{(b)}$Facult\'{e} des Sciences, Universit\'{e} Ibn-Tofail, K\'{e}nitra;$^{(c)}$Facult\'e des Sciences Semlalia, Universit\'e Cadi Ayyad, LPHEA-Marrakech;$^{(d)}$LPMR, Facult\'e des Sciences, Universit\'e Mohamed Premier, Oujda;$^{(e)}$Facult\'e des sciences, Universit\'e Mohammed V, Rabat;$^{(f)}$Institute of Applied Physics, Mohammed VI Polytechnic University, Ben Guerir; Morocco.\\
% ...
$^{45}$Physics Department, Southern Methodist University, Dallas TX; United States of America.\\
% ...
$^{49}$Fakult\"{a}t Physik , Technische Universit{\"a}t Dortmund, Dortmund; Germany.\\
% ...
$^{55}$II. Physikalisches Institut, Georg-August-Universit\"{a}t G\"ottingen, G\"ottingen; Germany.\\
% ...
$^{104}$CPPM, Aix-Marseille Universit\'e, CNRS/IN2P3, Marseille; France.\\
% ...
$^{119}$$^{(a)}$New York University Abu Dhabi, Abu Dhabi;$^{(b)}$United Arab Emirates University, Al Ain; United Arab Emirates.\\
% ...
$^{123}$Homer L. Dodge Department of Physics and Astronomy, University of Oklahoma, Norman OK; United States of America.\\
% ...
$^{157}$Raymond and Beverly Sackler School of Physics and Astronomy, Tel Aviv University, Tel Aviv; Israel.\\
% ...
$^{a}$ Also at Affiliated with an institute formerly covered by a cooperation agreement with CERN.\\
$^{b}$ Also at An-Najah National University, Nablus; Palestine.\\
$^{c}$ Also at Borough of Manhattan Community College, City University of New York, New York NY; United States of America.\\
$^{d}$ Also at Center for Interdisciplinary Research and Innovation (CIRI-AUTH), Thessaloniki; Greece.\\
$^{e}$ Also at Centre of Physics of the Universities of Minho and Porto (CF-UM-UP); Portugal.\\
$^{f}$ Also at CERN, Geneva; Switzerland.\\
% ...

\end{flushleft}


%

%
%

%

%
%
%
%

\end{document}